\documentclass{aa}  
\usepackage[hidelinks]{hyperref}
\hypersetup{
colorlinks=true,
linkcolor=blue,
citecolor = blue,
urlcolor=cyan}
\usepackage{graphicx}
\usepackage{subcaption}
\usepackage{aadefs}
\usepackage{txfonts}
\usepackage{dblfloatfix}  

\usepackage{graphicx}
\usepackage{txfonts}
\usepackage{lipsum}
\usepackage{subcaption}         
\usepackage{lscape}             
\usepackage{placeins}           
                                
\begin{document}

\title{High-resolution optical spectroscopy reveals the distinctive volatile composition of the interstellar comet 3I/ATLAS}
   \authorrunning{Aravind et al.}
    \titlerunning{Interstellar comet 3I/ATLAS}

    \author{K. Aravind\inst{1},
    E. Jehin\inst{1},
    C. Opitom\inst{2},
    J. Manfroid\inst{1},
    D. Hutsemékers\inst{1},
    D. Bodewits\inst{3},
    M. Bannister\inst{4},
    R. C. Dorsey\inst{5},
    L. Ferellec\inst{6},
    H. Kawakita\inst{7},
    M. Lippi\inst{8},
    C. Snodgrass\inst{2} 
    } 
    
    \institute{ 
    STAR Institute, Université de Liège, Allée du 6 Août 19c, 4000 Liège, Belgium\\\ \email{aravind.krishnakumar@uliege.be} \and Institute for Astronomy, University of Edinburgh, Royal Observatory, Edinburgh EH9 3HJ, UK \and Physics Department, Edmund C. Leach Science Center, Auburn University, Auburn, AL 36849, USA \and School of Physical and Chemical Sciences -- Te Kura Matū, University of Canterbury, Christchurch 8140, Aotearoa New Zealand \and Department of Physics, P.O. Box 64, 00014 University of Helsinki, Helsinki, Finland \and Faculty of Science and Engineering, Northumbria University, Newcastle NE1 8ST, UK \and Koyama Space Science Institute, Kyoto Sangyo University, Kamigamo Motoyama, Kita-ku, Kyoto 603-8555, Japan \and INAF -- Osservatorio Astrofisico di Arcetri, Largo Enrico Fermi 5, 50125 Firenze, Italy }

   \date{Received ; }

  \abstract
   {Comets preserve volatile material from the early stages of planetary-system formation. The discovery of the interstellar comet 3I/ATLAS enables a direct comparison between the composition of a comet formed around another star and that of comets in the Solar System.} 
   { We aim to characterise the evolution of the coma composition of 3I/ATLAS across perihelion by deriving the production rates of the key volatile species in the optical range, and to compare its composition with that of Solar System comets.}
   {We present a homogeneous high-resolution spectroscopic monitoring in the near-UV and the optical with UVES on the ESO Very Large Telescope, from August 2025 (r$_h$=3.14 au) pre-perihelion to February 2026 (r$_h$=4.34 au) post-perihelion. From the flux-calibrated spectra, we measured the gas emissions of OH, NH, CN, C$_3$, CH and C$_2$, derived their production rates, and examined abundance ratios and heliocentric-distance trends. We report an upper limit of the NH$_2$ production rate and also compare the OH-based water production with estimates from the forbidden oxygen lines.}
   {The gas production rates exhibit broadly symmetric pre- and post-perihelion behaviour, although their heliocentric-distance dependences are steeper than those typically observed in Solar System comets. NH and NH$_2$ are strongly depleted, indicating a severe deficiency of ammonia-related volatiles. The C$_2$/CN ratio points to a depleted composition and it evolves with heliocentric distance, consistent with coma-driven effects like hyperactivity rather than intrinsic compositional changes. Water production derived from forbidden oxygen lines exceeds OH-based estimates except near perihelion, suggesting varying contributions from H$_2$O and CO$_2$ to the coma chemistry. In relative abundance correlations, 3I/ATLAS departs from the canonical Solar System comet cluster and lies closest to the population of strongly carbon-chain depleted comets, including the G-Z-type comets, while exhibiting a strong depletion of NH- and C$_3$-bearing species relative to CN.}
   {3I/ATLAS extends the known diversity of cometary compositions beyond that observed in the Solar System, providing direct evidence that volatile reservoirs in other planetary systems can differ substantially from those sampled by Solar System comets.}

   \keywords{comets: general – comets: individual: Comet 3I/ATLAS – techniques: spectroscopic
               }

   \maketitle
\nolinenumbers

\section{Introduction}

Interstellar objects (ISOs) travelling through the Solar System give us a rare opportunity to study the properties of planetesimals that formed around other stars. Unlike Solar system comets, which represent the conditions of one protoplanetary disk, ISOs provide a unique sampling of various formation environments and evolutionary paths. Observations of these objects can then help to understand the commonality of comet compositions, volatile inventories, activity and ejection mechanisms across different planetary systems \citep{comets_exo,interstellar_interloppers}. 

The discovery of 1I/2017 U1 (‘Oumuamua) was the first confirmed identification of an ISO, but the lack of cometary activity limited what could be learned about its composition \citep{Meech2017, Jewitt2017,ye17-oumuamua, fitzsimmons2018}. The second ISO, 2I/Borisov \citep{Borisov2019}, was discovered before perihelion and exhibited clear cometary activity \citep{Jewitt2019BorisovNature}, enabling an extensive observational campaign. Dust studies revealed coma morphology and dust properties broadly comparable to active Solar System comets, although temporal variability and heterogeneous dust release were also reported \citep{jewitt2019,cremonese20202I,hui_borisov,guzik_borisov,yang_waterice}. Spectroscopic observations detected CN, C$_2$, NH$_2$, and water-related products, indicating that much of its volatile composition resembled that of Solar System comets \citep{borisov_aravind, opitom_borisov, Mckay_borisov_oxygen, lin_borisov,kareta_borisov, fitzsimmons2019}. However, sub-millimetre and ultraviolet observations revealed an unusually high CO abundance relative to water \citep{borisov_cordiner_CO, borisov_COrich}, making 2I/Borisov one of the most CO-rich comets known and suggesting formation beyond the CO snow line. Atomic nickel and iron were also detected in its coma \citep{guzik2021Nature, Borisov_highres}. Collectively, these observations demonstrated that while interstellar comets share many properties with Solar System comets, they can also exhibit distinctive compositional signatures reflecting different formation and evolutionary histories.

The discovery of comet 3I/ATLAS (C/2025 N1) in July 2025 as an active body—rather than an asteroid—was confirmed by the development of a dust coma and activity profile, consistent with the sublimation of volatiles as it approached the Sun \citep{seligman2025, Jewitt2025ATLAS}. Its discovery triggered an unprecedented international observing campaign spanning ground- and space-based spectroscopic, photometric, and polarimetric observations across multiple wavelengths.



Initial observations established 3I/ATLAS as an active, dust-rich comet with a distinctly red colour, redder than 2I/Borisov and comparable to the most dust-rich Solar System comets \citep{Bolin2025ATLAS}. Early optical spectroscopy at 4.47 au revealed a red, featureless continuum with no detectable molecular emissions, consistent with expectations for Solar System comets at similar heliocentric distances \citep{Opitom2025ATLAS}, while complementary observations measured a similar spectral slope and a rotation period of 16.79$\pm$0.23 h \citep{3I_10m}. Continued photometric monitoring showed increasing dust production, coma reddening, and steadily rising activity toward perihelion without evidence for fragmentation \citep{SantanaRos2025ATLAS,Jewitt2025ATLAS}.

The spectroscopic evolution of 3I/ATLAS has been particularly interesting and unique. Spectroscopic monitoring revealed the onset of atomic NiI emission and the first detection of CN while the comet was still at 3.65 au \citep{Rahatgaonkar2025UVES,SalazarManzano2025CN,hoogendam3I_Ni}, while near-infrared observations detected water ice mixed predominantly with amorphous carbon \citep{Yang2025NIR}. Subsequent high-resolution UVES monitoring (our dataset) traced the evolution of the NiI/FeI abundance ratio from extreme values toward those typical of Solar System comets \citep{Hutsemekers2026NiFe,3I_Damien_post}. This study was also complemented by post-perihelion observations from Keck/KCWI \citep{hoogendam2026keck} and low-resolution observation from 2 m class telescopes Lijiang and Xinglong  \citep{Zhao_3I_postper}.

\cite{hoogendam2026keck} reported a carbon-depleted composition for the comet with a C$_2$/CN ratio slightly higher than the upper limits reported during pre-perihelion observations \citep{SalazarManzano2025CN, 3I_lazzarin_3I}. However, subsequent post-perihelion observations using TRAPPIST narrow-band photometry and low-resolution spectroscopy suggested a more carbon-typical composition \citep{jehin2025a,jehin2025b,Zhao_3I_postper,3I_kawakita2026}. Water production rates measured before and after perihelion by multiple facilities provided valuable constraints on the evolution of the comet and the possible presence of an extended source \citep{3I_juncen_h2O, 3I_hanjie_h2o, 3I_Xing_h2O, 3I_combi_h2o, 3I_Lisse_h2o, 3I_Lisse_ApJ, 3I_lisse_2025Spherex}. Complementary studies have extended to polarimetric observations, revealing a complex interplay of dust, volatile release, and surface evolution \citep{Zuri2025Polarisation,2025RNAAS,zubko2025polarisation}. Throughout its observed passage through the Solar System, 3I/ATLAS exhibited no evidence of disintegration.

In parallel, the James Webb Space Telescope (JWST) has also played a pivotal role in advancing our understanding of 3I/ATLAS. Observations of the comet at a heliocentric distance of 3.32 au reporting high-sensitivity infrared spectra of the comet, revealed a CO$_2$ dominated gas coma \citep{JWST2025CO2}. The post-perihelion JWST observations in December 2025 at a heliocentric distance of 2.2 - 2.5 au, provide a detailed view of its volatile inventory, including the detection of methane gas \citep{JWST2026, 3I_JWST_Nathan}. Additionally, sub-mm observations have provided further insights into the gas and nucleus of 3I/ATLAS \citep{3I_Biver}. Pre-perihelion ALMA observation campaigns spanning heliocentric distances of 2.6–1.7 au detected enriched methanol abundance with respect to HCN \citep{3IRadio2025} and additionally \cite{3I_Biver} reports similar enhancement very close to perihelion. \cite{3I_Cyrielle_isotope}, making use of the best UVES/VLT spectra from our dataset, report $^{12}$C/$^{13}$C and $^{14}$N/$^{15}$N isotopic ratios for CN, \cite{3I_JWST_cordiner_2026}, utilising JWST observations at 2.4~au, report $^{12}$C/$^{13}$C for CO$_2$ and CO, and \cite{D_H_Nathan}, with ALMA observations, report D/H. All three studies find isotopic ratios higher than Solar System comets, pointing to a cold and distant origin of the interstellar comet.

Together, these studies have established a robust, multiwavelength baseline for understanding 3I/ATLAS’s activity and evolution. The comet has challenged expectations set by previous interstellar visitors, with a dust-rich, actively evolving coma and a unique spectroscopic fingerprint. In this context, our work presents a comprehensive analysis of the complete UltaViolet-Visible (UV-Vis) high-resolution spectroscopic activity evolution of 3I/ATLAS, providing a homogeneous high-resolution view of the complete UV-Vis spectroscopic evolution of this unique interstellar visitor.
 \begin{figure}[h!]
  \centering
   \includegraphics[width=0.95\linewidth]{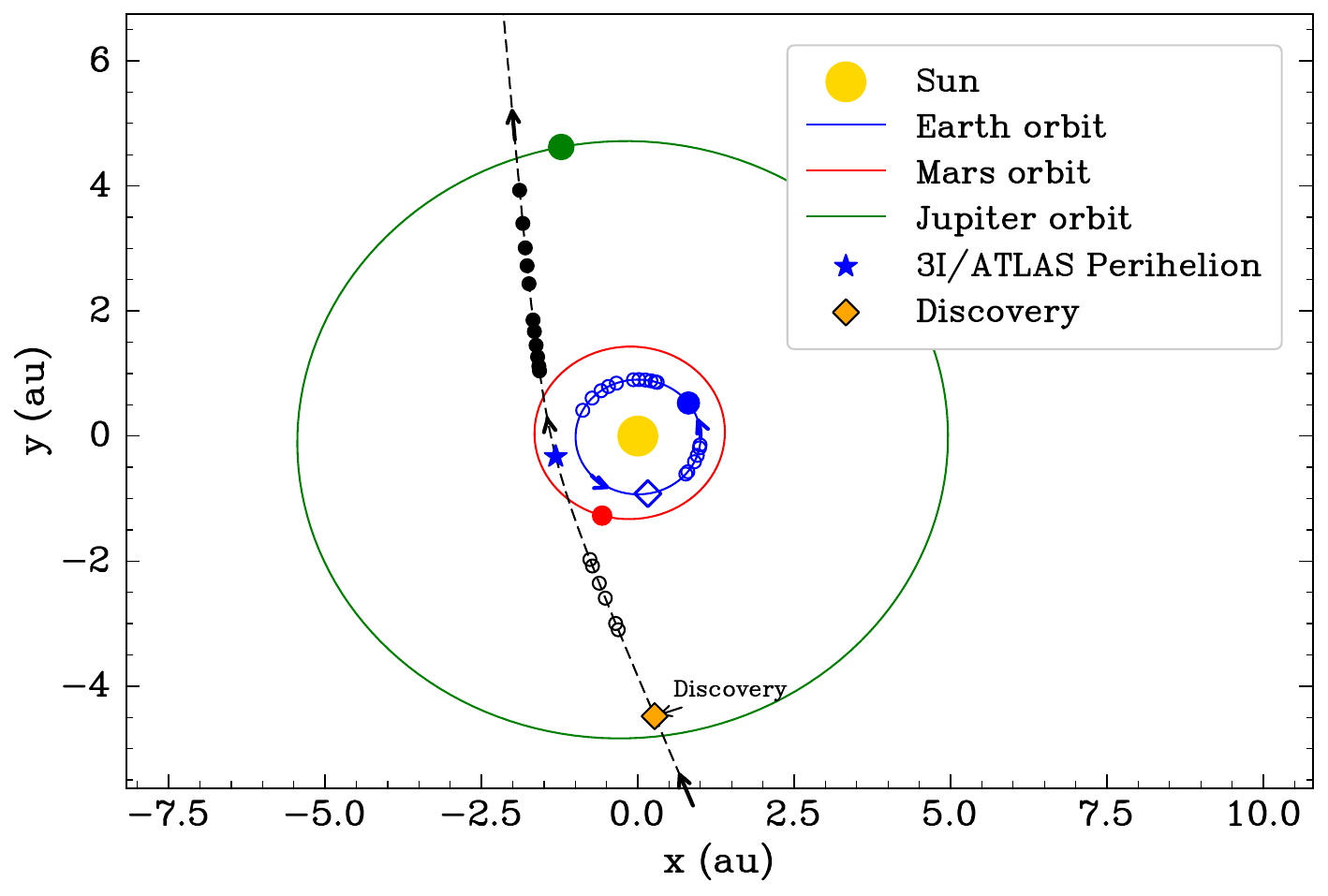}
      \caption{Orbit of 3I/ATLAS showing the 17 UVES/VLT observing epochs. Open and filled circles represent pre- and post-perihelion observations, respectively. Planet positions with filled circles are shown for the date of perihelion. The open diamond marks Earth's position at the discovery of 3I/ATLAS, while the open circles along Earth's orbit indicate its position at each observing epoch.
              }
         \label{orbit_obs}
   \end{figure}

\section{Observation and data reduction}

\begin{figure*}
\begin{subfigure}{0.49\linewidth}
\centering
\includegraphics[width = 0.92\textwidth]{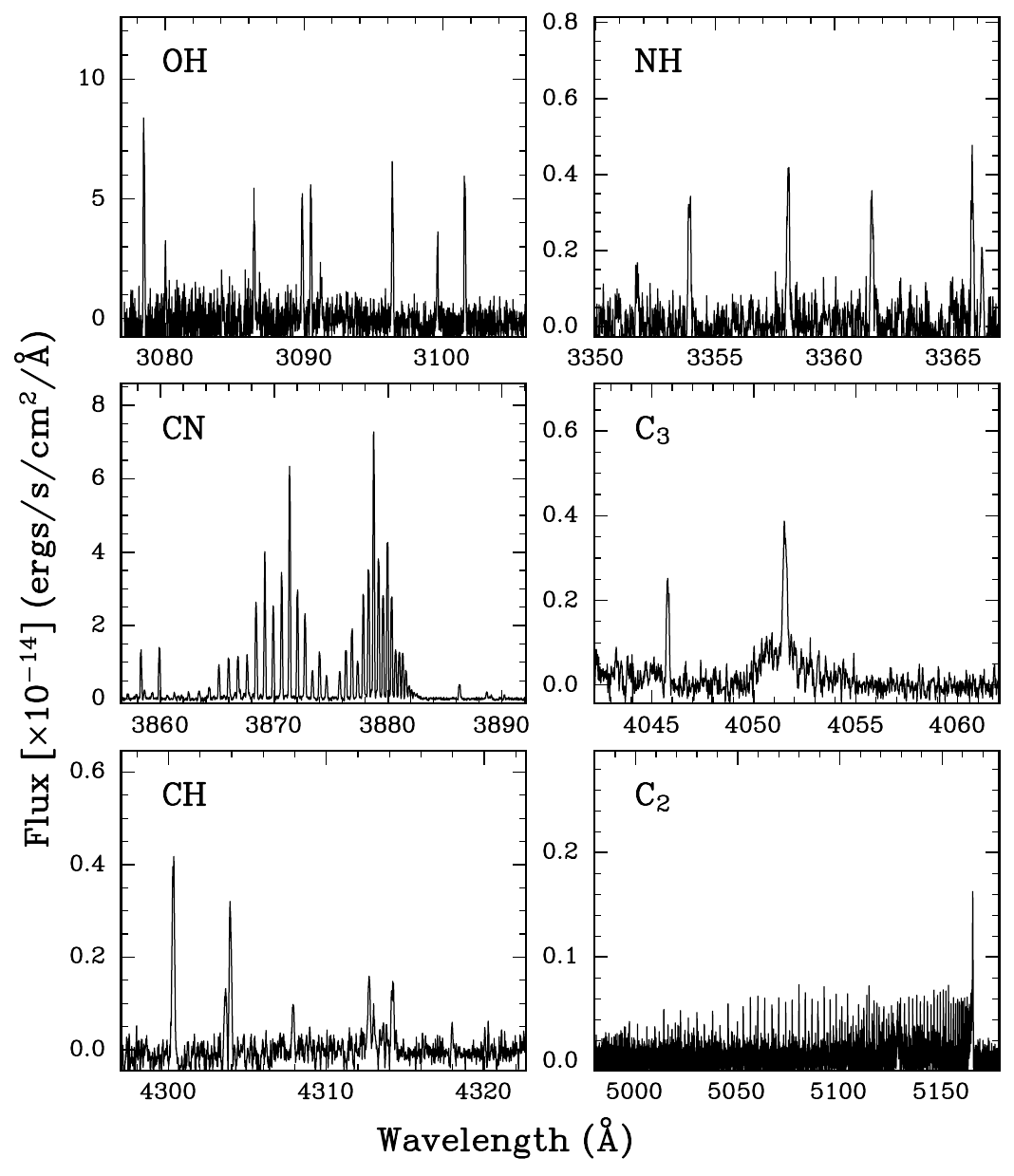}
\label{pre}
\end{subfigure}\hfill
\begin{subfigure}{0.49\linewidth}
\centering
\includegraphics[width = 0.92\textwidth]{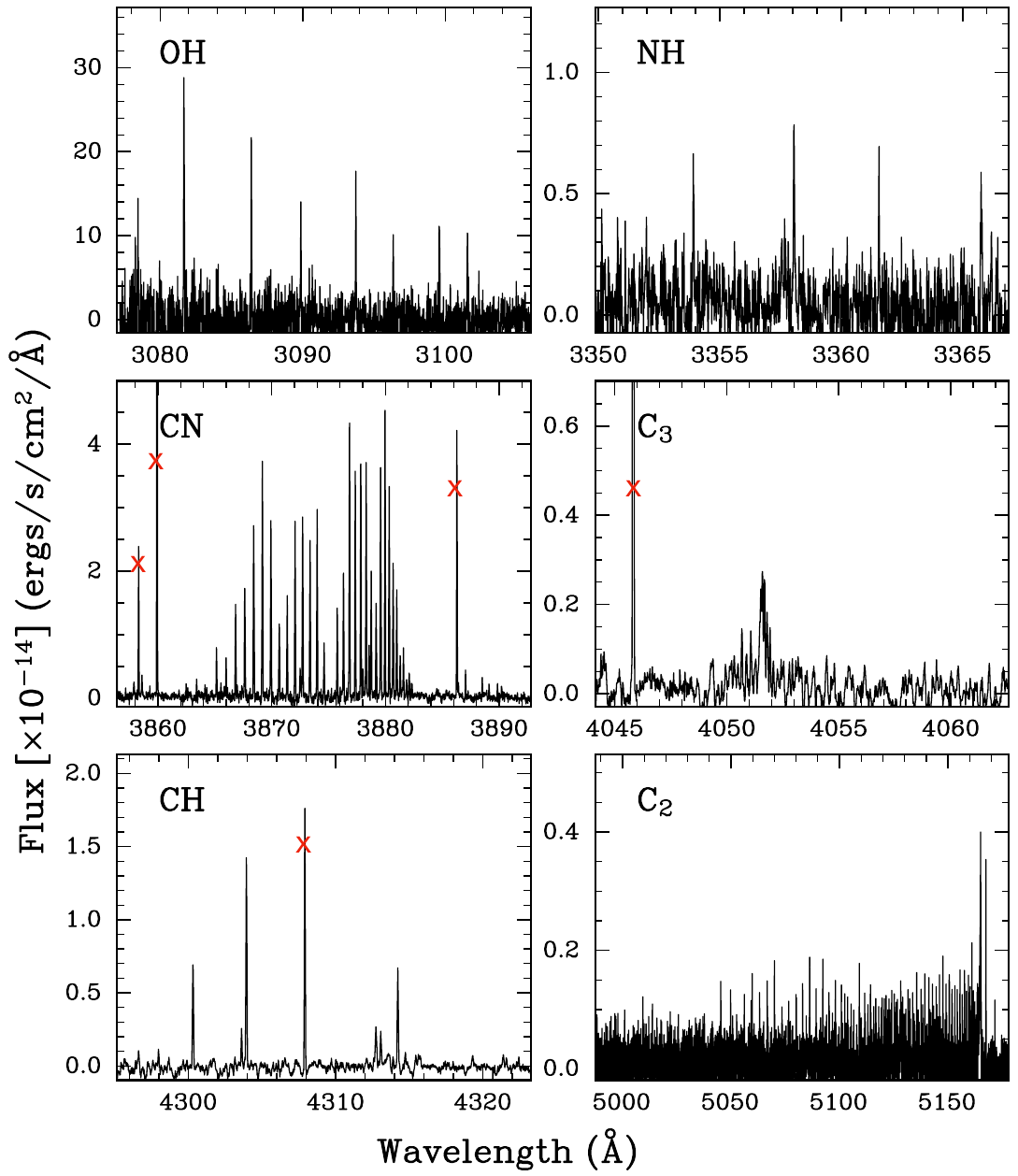}
\label{post}
\end{subfigure}\hfill
\caption{Emissions detected in 3I/ATLAS on epochs just before and after perihelion. The emission strengths on the y-axis are not a direct representation of the cometary activity, as the slit widths used in these observations are different. The red cross depicts the Ni\,I and Fe\,I emission lines. \textit{Left}: emissions detected in the UVES spectrum acquired before perihelion on 2025-09-14 at $r_h=2.14$ au. \textit{Right}: emissions detected in the UVES spectrum acquired after perihelion on 2025-12-06 at $r_h=1.93$ au.}
\label{pre_post}
\end{figure*}


\begin{table}[h!]
\caption{Observing circumstances of 3I/ATLAS on UVES}
\label{Obs_log}
\setlength{\tabcolsep}{6pt}
		\resizebox{\linewidth}{!}{%
\begin{tabular}{llllllll}
\hline
\hline
UT Date  & $r_{h}$ & $\dot{r}_{h}$  & $\Delta$  & $\dot{\Delta}$  & Settings & $w_{B} / w_{R}$ & $h_{B} / h_{R}$ \\
yyyy-mm-dd & (au) & $\mathrm{km}~\mathrm{s}^{-1}$ & (au) & $\mathrm{km}~\mathrm{s}^{-1}$ & nm & " & "\\\hline
 2025-08-12 & 3.14 & -55.3 & 2.69 & -15.4 & $346+580$ & 1.8/0.6 & 9.5/11.5 \\
 2025-08-15 & 3.04 & -54.9 & 2.66 & -13.3 & $390+580$ & 0.6/0.6 & 7.5/11.5 \\
 2025-08-28 & 2.64 & -52.9 & 2.59 & -6.60 & $348+580$ & 1.8/0.6 & 9.5/11.5 \\
 2025-08-28 & 2.64 & -52.9 & 2.59 & -6.60 & $437+860$ & 1.8/0.6 & 9.5/11.0 \\
 2025-09-03 & 2.46 & -51.6 & 2.57 & -4.80 & $348+580$ & 1.8/1.2 & 9.5/11.5 \\
 2025-09-03 & 2.46 & -51.6 & 2.57 & -4.80 & $437+860$ & 1.8/1.2 & 9.5/11.0 \\
 2025-09-04 & 2.43 & -51.3 & 2.57 & -4.60 & $348+580$ & 1.8/1.2 & 9.5/11.5 \\
 2025-09-04 & 2.43 & -51.3 & 2.57 & -4.60 & $437+860$ & 1.8/1.2 & 9.5/11.0 \\
 2025-09-10 & 2.25 & -49.5 & 2.55 & -3.70 & $348+580$ & 1.8/1.2 & 9.5/11.5 \\
 2025-09-11 & 2.22 & -49.2 & 2.55 & -3.70 & $437+860$ & 1.8/1.2 & 9.5/11.0 \\
 2025-09-12 & 2.19 & -48.8 & 2.55 & -3.60 & $348+580$ & 1.8/1.2 & 9.5/11.5 \\
 2025-09-14 & 2.14 & -48.1 & 2.54 & -3.60 & $437+860$ & 1.8/1.2 & 9.5/11.0 \\
 2025-12-04$^*$ & 1.88 & +43.3 & 1.88 & -17.04 & $~~~~~~~~~~580$& ~~~~~/0.6   & ~~~~~/11.5 \\
 2025-12-06$^\dag$ & 1.93 & +44.5 & 1.86 & -15.49 & $348+580$ & 0.6/0.6 & 9.5/11.5 \\
 2025-12-10 & 2.04 & +46.5 & 1.83 & -11.66 & $437+860$ & 0.6/0.6 & 9.5/11.0 \\
 2025-12-15 & 2.17 & +48.6 & 1.80 & -5.84 & $348+580$ & 0.6/0.6 & 9.5/11.5 \\
 2025-12-21 & 2.34 & +50.5 & 1.79 & +2.69  & $348+580$ & 0.6/0.6 & 9.5/11.5 \\
 2025-12-21 & 2.34 & +50.5 & 1.79 & +2.69 & $437+860$ & 0.6/0.6 & 9.5/11.0 \\
 2025-12-26 & 2.49 & +51.9 & 1.82 & +11.18 & $348+580$ & 0.6/0.6 & 9.5/11.5 \\
 2026-01-11 & 2.98 & +54.7 & 2.05 & +38.85 & $348+580$ & 1.2/1.2 & 9.5/11.5 \\
 2026-01-19 & 3.23 & +55.6 & 2.26 & +50.92 & $348+580$ & 1.2/1.2 & 9.5/11.5 \\
 2026-01-27 & 3.50 & +56.2 & 2.52 & +61.41 & $348+580$ & 1.8/1.8 & 9.5/11.5 \\
 2026-02-07 & 3.85 & +56.9 & 2.94 & +71.93 & $348+580$ & 1.8/1.8 & 9.5/11.5 \\
 2026-02-22 & 4.34 & +57.6 & 3.61 & +81.93 & $348+580$ & 1.8/1.8 & 9.5/11.5 \\
 \hline
\hline
\multicolumn{8}{p{1.3\linewidth}}{\textbf{Notes}. r$_h$ and $\Delta$ are the heliocentric and geocentric distances of the comet. $\dot{r}_{h}$ and $\dot{\Delta}$ are the corresponding velocities. w$_B$, w$_R$, h$_B$, and h$_R$ refer to the blue/red slit width and height, respectively; ($^*$): No observation with the setting 348 due to a technical problem; ($^\dag$): This observation was done
through light clouds.}
\end{tabular}}
\end{table}

Unlike the previous interstellar comet 2I/Borisov, 3I/ATLAS had a closer perihelion at 1.35 au and was observable at optical wavelengths for a longer period of time, except for a window around the perihelion when the comet was behind the Sun (see Figure \ref{orbit_obs}).
Observations took place between August 12, 2025 and February 22, 2026, using the Very Large Telescope (VLT) at the European Southern Observatory (ESO), with the UV-Visible Echelle Spectrograph (UVES \footnote{UVES User Manual, VLT-MAN-ESO-13200-1825,\\ \url{https://www.eso.org/sci/facilities/paranal/
instruments.html}}). Standard configurations 346+580 nm (dichroic 1), 390+580 nm (dichroic 1), and 437+860 nm (dichroic 2) were used, providing blue settings which cover spectral ranges of 3030-3880~\AA, 3260-4540~\AA, and 3730-4990~\AA, and red settings which cover spectral ranges of 4760-6840~\AA~and 6600-10600~\AA, respectively. Additionally, a non-standard 348 nm setting (covering 3050-3900~\AA) was used to simultaneously observe the OH (0-0) and CN (0-0) bands. Depending on the comet brightness, the slit width in the blue and red were varied between 1.8\arcsec and 0.6\arcsec~yielding a resolving power of roughly 35,000 - 65,000 respectively. A summary of the observing conditions is given in Table \ref{Obs_log}.

Raw frames were initially cleaned of cosmic ray hits using the Python implementation of the “lacosmic” package \citep{LAcosmic,lacosmic2012}. The UVES pipeline \citep{uves_pipeline} within the EsoReflex software \citep{esoreflex}   was then used for the reduction process, resulting in wavelength- and flux-calibrated two-dimensional spectra. The absolute flux calibration was carried out either using the archived master response curve (for standard settings) or a response curve derived from a standard star observed on the same night as the science spectrum (for the non-standard setting 348 nm). Final one-dimensional spectra were extracted by integrating over the full slit length (see Table \ref{Obs_log}), using IRAF modules.

The emission spectrum contains a continuum which is mostly made up of  dust-reflected sunlight and, in some cases, the twilight
contamination and the lunar sky background. To remove this underlying continuum, a solar reference spectrum \citep{kurucz_solar} adjusted for each of these components, following the procedure described by \cite{Manfroid2009} was adopted. This method accounts for the different Doppler shifts arising from the relative motions between the observer, the Moon, the Sun, and the comet—a necessity at the spectral resolution of UVES. Specifically the lunar sky background exhibits a radial velocity equal to the sum of the heliocentric and topocentric velocities of the Moon. The dust-reflected solar continuum is shifted by the combined heliocentric and topocentric velocities of the comet. The twilight component is shifted according to the topocentric velocity of the Sun, while the cometary emission itself is shifted by the geocentric velocity of the comet. These corrections ensure accurate subtraction of the solar continuum from the comet spectra providing a spectra consisting of only the gas component. The relevant velocity informations were retrieved from the JPL Horizon ephemeris generator.

\section{Data analysis and results}
\subsection{Overview of detected emissions}
The UVES monitoring started on August 12, 2025 when the comet was at a heliocentric distance of 3.14 au and it reveals a clear evolution in the molecular inventory of the coma. During the earliest epochs, the spectrum was dominated by CN ($B^{2}\Sigma^{+} - X^{2}\Sigma^{+}\;(0,0)$) together with strong atomic NiI and FeI emission \citep{Rahatgaonkar2025UVES, Hutsemekers2026NiFe}, while OH ($A^{2}\Sigma^{+} - X^{2}\Pi_{i}\;(0,0)$), NH ($A^{3}\Pi_{i} - X^{3}\Sigma^{-}\;(0,0)$), C$_3$ ($\tilde{A}^{1}\Pi_{u} - \tilde{X}^{1}\Sigma_{g}^{+}$), CH ($A^{2}\Delta - X^{2}\Pi\;(0,0)$) and C$_2$ ($d^{3}\Pi_{g} - a^{3}\Pi_{u}\;(0,0)$) became progressively detectable as the comet moved closer to the Sun. The doppler shifted red forbidden oxygen line (hereafter [OI]) at 6300\,\AA~has also been detected on several epochs. No ionic lines were detected at any of the epochs, probably due to the small slit size centered on the inner coma. The comet was not observable between 2025-09-15 and 2025-12-03, as its orbital position kept it below an elevation of 20$^\circ$ for Paranal throughout this period (see Figure \ref{orbit_obs}). The comet was further followed intensively until February 22, 2026, when all the emissions were seen to subside. This provides us with a large data set, totalling to 17 epochs of observations with regular monitoring at every $\sim$0.35 au across perihelion.

The UVES dataset is particularly valuable because it provides a single-instrument, homogeneous time series spanning both inbound and outbound evolution. This avoids systematic differences introduced by comparing production rates derived from different slit sizes, spectral resolutions, and reduction pipelines. The observed sequence therefore offers a robust basis for comparing compositional evolution, heliocentric scaling, and abundance ratios across perihelion. 

\subsection{Production rate measurements}
Emissions from OH, NH, CN, C$_3$, CH and C$_2$ were clearly detected before and after perihelion (see Figure \ref{pre_post}). No emission from NH$_2$ is observed during any of the epochs as noted also by \cite{3I_kawakita2026}. Since the spectrum of this comet was flooded with strong emissions from NiI and FeI, special care was given to select the lines corresponding to a given species while computing the respective production rates. The cometary line atlas developed at the University of Liege was used for this purpose \citep{Hardy_atlas, jehin_atlas_epsc}\footnote{\url{https://www2.cometa.uliege.be/Cometary_Lineatlas/Cometary_atlas.html}}.

In order to compute the production rates from the spectroscopic data via Haser coma outflow model \citep{haser57}, the full slit spectral extractions were used along with the equation,
\begin{equation}\label{prod_rate_eq}
    Q = \frac{4 \pi \Delta^2 v_{out}}{g~l_d}\times Flux \times HC,
\end{equation}
where $\Delta$ is the geocentric distance in au, g is the fluorescence efficiency (ergs/molecules/s), v$_{out}$ is the outflow velocity assumed to be 0.85$\times{r_h}^{-0.5}$ km/s \citep{cochran_OH_H2O}, where r$_h$ is the heliocentric distance and l$_d$ is the scale length of the daughter molecule. Several studies of comet 3I have reported HCN expansion velocities that are lower than those predicted by the commonly adopted relation \citep{3I_Biver, 3I_JWST_cordiner_2026, coulson_2026, 3IRadio2025}. However, the velocity appropriate for a Haser-equivalent daughter distribution does not need to be identical to the parent-gas velocity, since the daughter molecule can acquire an additional velocity through photodissociation.  With the empirical scale-lengths used in this paper, the adopted velocity primarily affects the normalisation of the derived production rates, as the expansion velocity enters the conversion from the measured column density (N) to the production rate, giving a first-order dependence $Q\propto Nv_{out}$. As an independent comparison, \citet{Mura_CN}, using the same velocity relation and deriving a new set of CN scale-lengths, obtained a CN production rate in good agreement with our value for the same observing epoch. We  therefore retain the widely used scale-lengths and velocity relation for the daughter molecules to provide a homogeneous treatment with the Solar System comet comparison sample.
\begin{table}[htbp]
\centering
\caption{Haser model parent and daughter scale-lengths and fluorescence efficiencies ($g$-factors) at 1.0 au, used for the production rate calculations.}
\label{haser_parameters}
\setlength{\tabcolsep}{6pt}
		\resizebox{\linewidth}{!}{%
\begin{tabular}{lcccc}
\hline
Molecule & Parent & Daughter & $g$-factor & Reference \\
         & ($\times$10$^4$ km) & ($\times$10$^4$ km) & (erg/s/mol) & \\
\hline
OH (0,0)      & 2.4 & 16 & $1.49 \times 10^{-15}$ &1,2 \\
NH (0,0)      & 5.0 & 15 & $6.27 \times 10^{-14}$ & 1,3\\
CN ($\Delta v = 0$) & 1.3 & 21 & $2.62 \times 10^{-13}$ & 1,4 \\
C$_3$($\lambda = 4050$~\AA) & 0.28 & 2.7 & $1.00 \times 10^{-12}$ & 1 \\
CH(0-0) & 7.8 & 0.48 & $1.05 \times 10^{-13}$ & 5\\
C$_2$($\Delta v = 0$) & 2.2 & 6.6 & $4.50 \times 10^{-13}$ & 1\\
NH$_2$(0-8-0)& 0.34 & 28.5 & 1.24 $\times$ 10$^{-15}$ & 6\\
\hline
\multicolumn{5}{p{1\linewidth}}{1 : \cite{Ahearn_85}; 2 : \cite{OH_g}; 3 : \cite{NH_g}; 4 : \cite{schleicher_CN_2010}; 5 : \cite{cochran_30years}; 6 : \cite{kawakita_NH2_g}}
\end{tabular}}
\end{table}

\begin{table*}[h!] \caption{Measured fluxes and derived production rates. } 
\label{prod_rate} 
\setlength{\tabcolsep}{3pt} 
\resizebox{\linewidth}{!}{%
\begin{tabular}{lcc|ccccccc|ccccccc} 
\hline \hline 
UT Date & $r_h$ & $\Delta$ & \multicolumn{7}{c|}{Flux [$10^{-15}$ erg\,s$^{-1}$\,cm$^{-2}$]} & \multicolumn{7}{c}{Production rates, $\log_{10} Q$ (error) [molecules s$^{-1}$]} \\
& (au) & (au) & OH & H$_2$O [OI] & NH & CN & C$_3$ & CH & C$_2$ & OH & H$_2$O [OI] & NH & CN & C$_3$ & CH & C$_2$ \\ \hline
2025-08-12 & 3.14 & 2.69 & -- & 0.18 & -- & -- & -- & -- & -- & -- & 27.22(0.14) & -- & 24.2(0.30)$^*$ & -- & -- & -- \\ 
2025-08-15 & 3.04 & 2.66 & -- & 0.15 & -- & 1.06 & -- & -- & -- & -- & 27.12(0.14) & -- & 24.48(0.05) & -- & -- & -- \\
2025-08-27 & 2.64 & 2.59 & 9.34 & -- & -- & 15.1 & 0.46 & $<0.12$ & 5.69 & 26.60(0.01) & -- & -- & 24.85(0.01) & 22.53(0.33) & $<24.11$ & 24.42(0.26) \\
2025-09-03+04$^a$ & 2.43 & 2.57 & 15.0 & -- & 0.36 & 29.5 & 1.14 & 0.45 & 5.64 & 26.97(0.00) & -- & 24.16(0.18) & 25.02(0.01) & 22.80(0.41) & 24.63(0.21) & 24.44(0.26) \\
2025-09-10+11$^a$ & 2.23 & 2.55 & 45.0 & -- & 0.75 & 78.8 & 2.94 & 1.01 & 9.70 & 27.19(0.05) & -- & 24.36(0.13) & 25.29(0.01) & 23.07(0.25) & 24.92(0.06) & 24.56(0.01) \\
2025-09-12+14$^a$ & 2.16 & 2.55 & 48.8 & -- & 0.89 & 108 & 3.02 & 1.77 & 12.6 & 27.28(0.04) & -- & 24.41(0.18) & 25.37(0.01) & 23.10(0.15) & 25.11(0.03) & 24.63(0.01) \\
2025-12-04 & 1.88 & 1.88 & -- & 5.95 & -- & -- & -- & -- & 39.3 & -- & 28.14(0.19) & -- & -- & -- & -- & 25.21(0.01) \\ 
2025-12-06$^\dag$ & 1.93 & 1.86 & 101 & 3.43 & 0.98 & 46.3 & -- & -- & 26.0 & 27.68(0.03) & 28.03(0.14) & 24.78(0.08) & 25.34(0.01) & -- & -- & 25.09(0.01) \\ 
2025-12-10 & 2.04 & 1.83 & -- & -- & -- & 34.9 & 3.28 & 2.74 & -- & -- & -- & -- & 25.29(0.02) & 23.39(0.15) & 25.61(0.03) & -- \\
2025-12-15 & 2.17 & 1.80 & 55.4 & 2.04 & 0.45 & 21.5 & -- & -- & 12.9 & 27.49(0.03) & 27.86(0.15) & 24.53(0.10) & 25.13(0.02) & -- & -- & 24.83(0.03) \\
2025-12-21 & 2.34 & 1.79 & 23.9 & -- & 0.17 & 11.6 & 0.85 & 0.55 & 5.53 & 27.04(0.04) & -- & 24.30(0.14) & 24.97(0.03) & 22.96(0.30) & 25.13(0.05) & 24.63(0.03) \\
2025-12-26 & 2.49 & 1.82 & 10.6 & 0.72 & 0.16 & 7.09 & -- & -- & 4.98 & 26.93(0.03) & 27.51(0.14) & 24.18(0.16) & 24.81(0.04) & -- & -- & 24.44(0.20) \\
2026-01-11 & 2.98 & 2.05 & 9.47 & 0.35 & -- & 3.36 & -- & -- & $<1.99$ & 26.21(0.17) & 27.12(0.14) & -- & 24.36(0.08) & -- & -- & $<23.80$ \\ 2026-01-19 & 3.23 & 2.26 & $<4.29$ & 0.28 & -- & 1.06 & -- & -- & $<0.95$ & $<26.07$ & 27.12(0.15) & -- & 24.06(0.15) & -- & -- & $<23.12$ \\ 2026-01-27 & 3.50 & 2.50 & $<2.77$ & 0.27 & -- & 0.77 & -- & -- & $<0.88$ & $<25.83$ & 27.10(0.19) & -- & 23.68(0.26) & -- & -- & $<22.80$ \\
2026-02-07 & 3.85 & 2.94 & -- & -- & -- & 0.32 & -- & -- & -- & -- & -- & -- & 23.56(0.32) & -- & -- & -- \\ \hline
\hline
\multicolumn{17}{p{0.8\linewidth}}{$^*$ Adopted from \cite{Hutsemekers2026NiFe}}\\
\multicolumn{17}{p{1.3\linewidth}}{$^a$ Spectra obtained with complementary UVES settings on consecutive nights were combined to achieve complete wavelength coverage. The production rates were derived from the resulting merged spectrum.}\\
\multicolumn{17}{p{0.8\linewidth}}{$^\dag$ The production rates are slightly underestimated due to atmospheric extinction.}\\
\multicolumn{17}{p{1.3\linewidth}}{\textbf{Notes.} Fluxes are given in units of $10^{-15}$ erg\,s$^{-1}$\,cm$^{-2}$ and Production rates are given as $\log Q$ in molecules s$^{-1}$, with uncertainties in parentheses.}\\
\multicolumn{17}{p{0.8\linewidth}}{ The reported uncertainties correspond to 1$\sigma$ absolute errors.}
\end{tabular}}
\end{table*}

Parent and daughter scale-lengths of OH, NH, CN, C$_2$ and C$_3$ were adopted from \cite{Ahearn_85} and are then scaled to $r_h^{2}$. Although the fluorescence efficiencies of C$_2$ and C$_3$ have been taken from \cite{Ahearn_85} and scaled to $r_h^{-2}$, that of OH \citep{OH_g}, NH \citep{NH_g} and CN \citep{schleicher_CN_2010} have been computed based on the heliocentric distances and velocities and then scaled to $r_h^{-2}$. The scale-lengths and fluorescence efficiency of CH was taken from \cite{cochran_30years}. The parameters required to compute the upper limit of NH$_2$ production rates were adopted from \cite{kawakita_NH2_g}. See Table \ref{haser_parameters} for the values of scale length and fluorescence efficiency at 1.0 au. Flux in Equation \ref{prod_rate_eq} is the integrated emission-band flux of the molecule of interest, and HC is the Haser correction, defined as the inverse of the Haser factor \citep{Fink1996}. The error in flux is computed from the continuum region
close to the bandpass of the molecule of interest and propagated further to obtain the error in the production rates. The Haser correction corresponding to each molecule is computed for the slit area (see Table \ref{Obs_log}) as explained in \cite{aravind_2020F3}.

\begin{figure}[h!]
 \centering
  \includegraphics[width=0.95\linewidth]{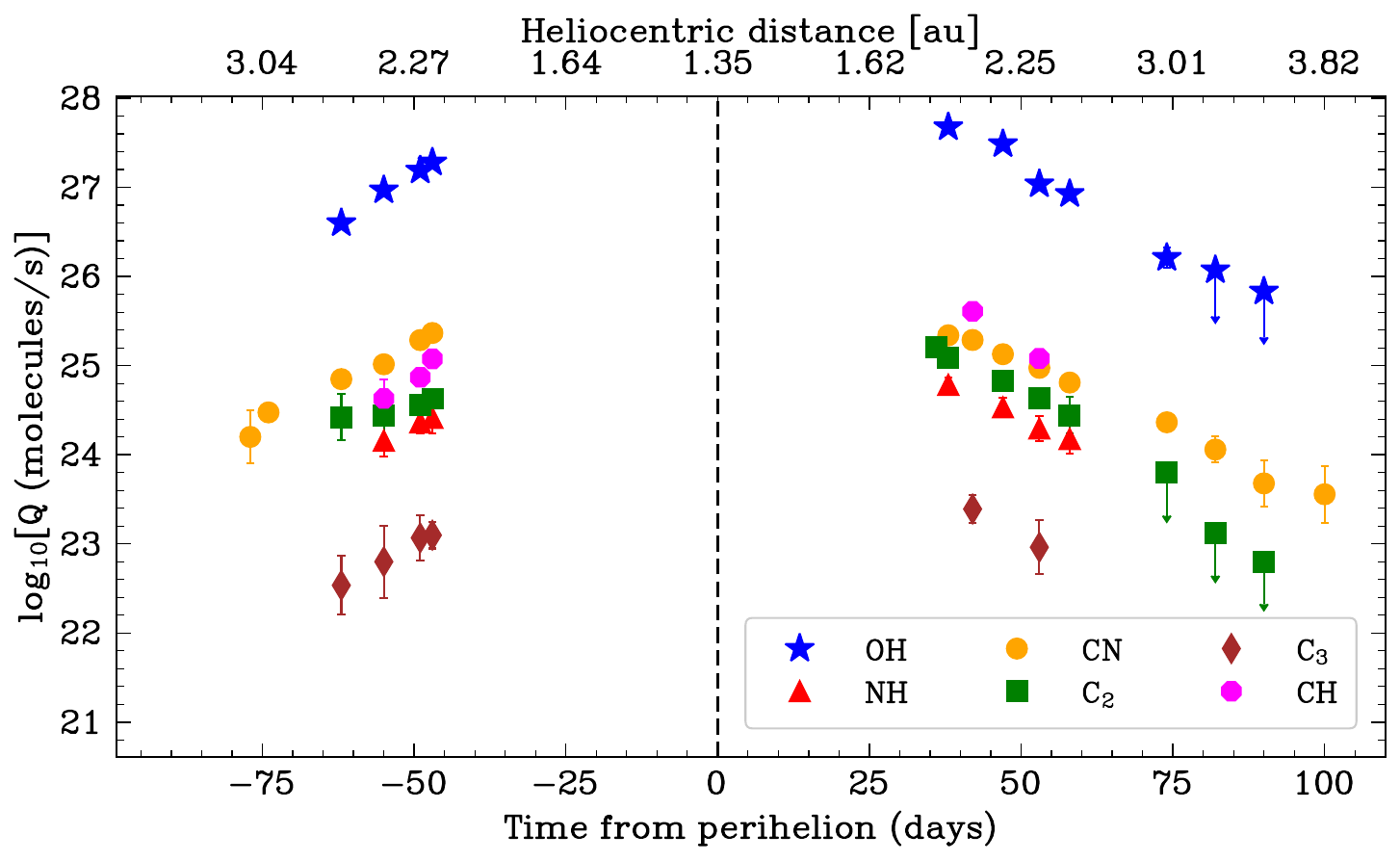}
     \caption{The gas production rates of comet 3I/ATLAS as a function of days to perihelion and heliocentric distance. The error bars correspond to 1$\sigma$ absolute errors.
             }
        \label{Qrates}
  \end{figure}

The computed production rates and their absolute errors are provided in Table \ref{prod_rate} and illustrated in Figure \ref{Qrates} as a function of days to perihelion and heliocentric distance ($r_h$). 
For the first epoch, the CN emission band is not fully covered by the 346 nm setting (see top left panel in Figure \ref{panel1}). We therefore adopt the CN production rate reported by \cite{Hutsemekers2026NiFe}, derived from the same dataset and using the same parameters as in this work, based on a model reconstruction of the complete CN band \citep{Manfroid2009}. For the last few epochs, 3-$\sigma$ upper limits are reported for emissions that are not detected. The production-rate sequence shows a steady rise in activity as 3I/ATLAS approached perihelion, followed by a decline after perihelion. OH, CN, CH and C$_2$ are all clearly detected in the data set, while there is only weak detection of NH and C$_3$ close to perihelion. The evolution of all the emission bands across the various epochs of observation are illustrated in Figures \ref{panel1} and \ref{panel2}. The weakness of the NH emission and the absence of any NH$_2$ lines, even close to perihelion, along with the possibly N$_2$ rich composition \citep{Lea_3I}, indicates a depletion in certain nitrogen-bearing species in the coma. On the other hand, the first active interstellar comet 2I/Borisov was rich in NH$_2$ \citep{Deam_2026}. With the help of fluorescence modelling techniques described in \cite{Kawakita_2001_NH2}, a synthetic spectrum of the NH$_2$ (0-8-0) band, corresponding to the production rate of NH on 2026-12-06, was used to compute log$_{10}$Q(NH$_2$) = 24.49 molec/s. The production rate was computed from the synthetic spectrum flux using Equation \ref{prod_rate_eq} with NH$_2$ scale-lengths and fluorescence efficiencies from \cite{kawakita_NH2_g}. The modeled NH$_2$ emission lines are not distinguishable from the noise in the observed spectrum.
Hence, the above mentioned production rate is being reported as the upper limit for NH$_2$ in 3I/ATLAS.

The CN production rate is the best-sampled molecular tracer in the data set and provides the clearest view of the comet’s gas evolution. CN strengthens steadily as the comet approaches perihelion, and its evolution broadly follows a power law in heliocentric distance (see Section \ref{powerlaw_sec}). C$_2$ activity was observed to be varying steeply with respect to heliocentric distance and had diminished significantly by 11 January 2026 when the comet was at a heliocentric distance of 2.98 au.

We also computed the water production rate using the integrated band flux of [OI] at 6300\,\AA. This approach has previously been used in many cases \citep[eg., ][]{schultz_1992_oxygen, Fink1996, Morgenthaler_2001,Morgenthaler_2007, Mckay_2012_oxygen, 67P_muse_oxygen, Mckay_borisov_oxygen, aravind_2020F3, Shinnaka_3I_oxygen}. Thanks to the high-resolution mode, the Doppler shifted 6300\,\AA~ [OI] line is free of any [OI] sky contamination. We follow the method described in \cite{aravind_2020F3} without the requirement of decontamination from NH$_2$ emissions as they are not detected. The derived water production rates and their comparison with the OH band are detailed in Section \ref{OH_O1} and reported in Table \ref{prod_rate}. 


\subsection{Heliocentric-distance scaling - Power law}\label{powerlaw_sec}

\begin{figure*}[h!]
  \centering
\includegraphics[width=0.90\linewidth]{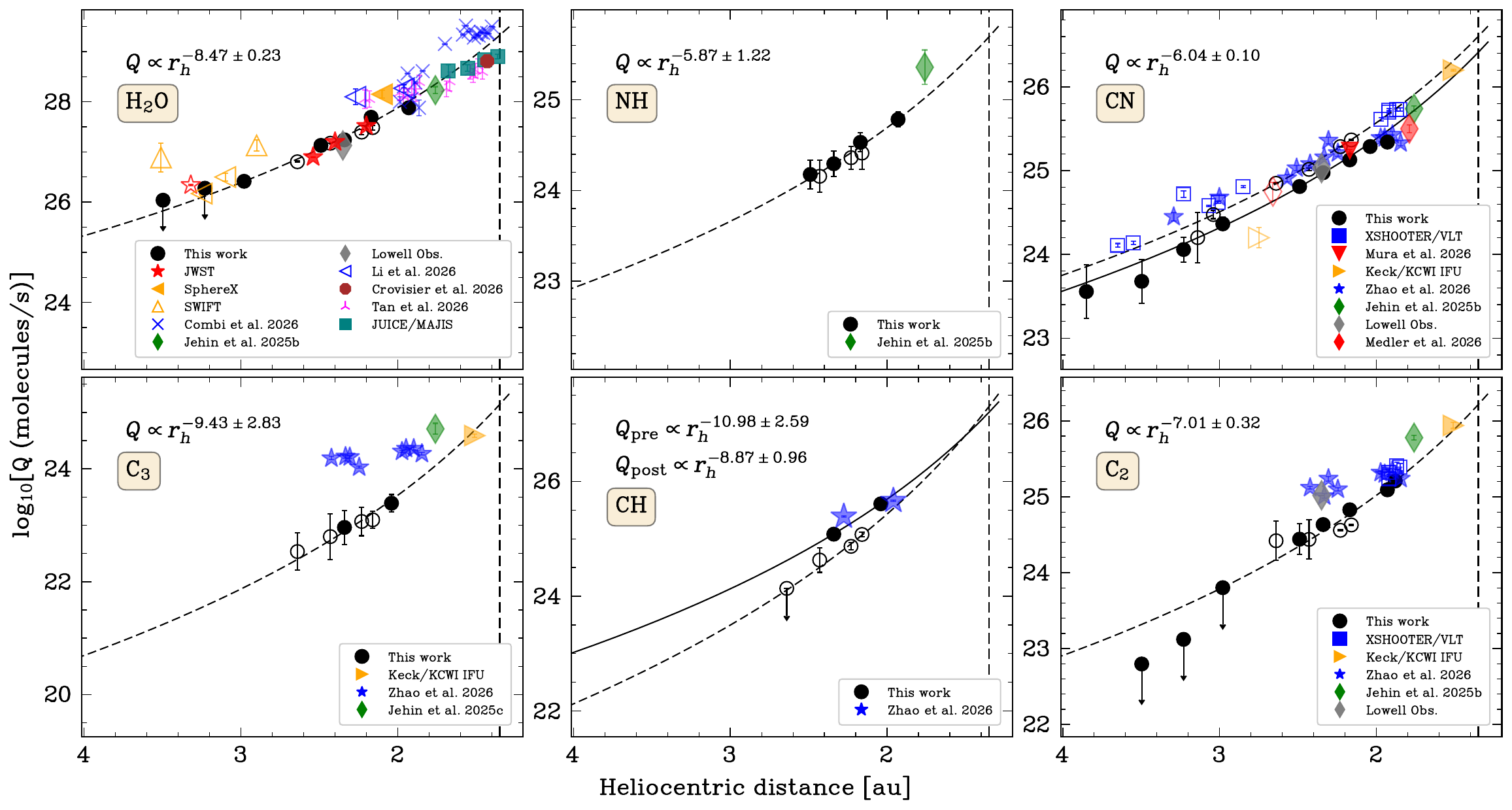}
\caption{The power law fitting the various production rates of comet 3I/ATLAS computed in this work overlaid with production rates reported by JWST \citep{JWST2026,JWST2025CO2,3I_JWST_Nathan}, SphereX \citep{3I_lisse_2025Spherex, 3I_Lisse_h2o}, SWIFT \cite{3I_Xing_h2O}, \cite{3I_combi_h2o}, TRAPPIST \citep{jehin2025a,jehin2025b}, \cite{3I_juncen_h2O}, \cite{3I_hanjie_h2o}, XSHOOTER/VLT \citep{Rahatgaonkar2025UVES}, KeckII/KCWI IFU \citep{hoogendam3I_Ni,hoogendam2026keck}, JUICE/MAJIS \citep{3I_JUICE}, \cite{3I_medler_hoogendam}, \cite{Zhao_3I_postper}, \cite{Mura_CN} and Lowell Observatory (David Schleicher - Private communication). Open symbols are for pre-perihelion measurements and filled for post-perihelion. The error bars correspond to 1$\sigma$ absolute errors.}
\label{powerlaw}
\end{figure*}

For each major species detected, we fitted a power law of the form $Q\propto r_h^{-n}$. To date, several studies have reported pre- and post-perihelion production rates of H$_2$O \citep{3I_combi_h2o,JWST2026, 3I_JWST_cordiner_2026, 3I_juncen_h2O, 3I_JWST_Nathan, JWST2025CO2, 3I_hanjie_h2o, 3I_Xing_h2O, 3I_lisse_2025Spherex, 3I_Lisse_h2o, 3I_Lisse_ApJ, 3I_JUICE}; OH \citep{jehin2025b, jehin2025a, Crovisier2025}; and CN, C$_2$, and C$_3$ \citep{Rahatgaonkar2025UVES, hoogendam3I_Ni, 3I_medler_hoogendam, hoogendam2026keck, 3I_lazzarin_3I, jehin2025b, jehin2025a, Zhao_3I_postper, Mura_CN}. NH has been reported only by \citep{jehin2025a} and \cite{Mura_CN} (upper limit), while the absence of NH$_2$ was noted by \cite{3I_kawakita2026}. The power law analysis is restricted to the results obtained from the UVES observations and the production rates from the various publications mentioned above were over plotted for comparison. For the purpose of uniformity, the OH production rates were converted to H$_2$O using the relation, Q(H$_2$O) = 1.36 $\times$ r$_h^{-0.5}~\times$ Q(OH) \citep{Schleicher1998}. The fits for H$_2$O, NH, CN, C$_3$, CH, and C$_2$ are illustrated in Figure \ref{powerlaw}. For uniformity, the production rates compared with have been scaled to the velocity used in this work. The black circles represent the UVES results in which open circles are for pre-perihelion measurements and filled circles are for post-perihelion measurements. 

Separate power-law fits to the pre- and post-perihelion data yield consistent slopes within the uncertainties for all the regular species detected with UVES. This demonstrates a symmetric activity trend about perihelion and motivates fitting the entire dataset of each specie with a single power law. Water production rates computed in this work followed a consistent power law  with a heliocentric index of $n = 8.47\pm0.23$. Meanwhile, it is to be noted that a few of the reported water production rates \citep{3I_juncen_h2O, 3I_Xing_h2O, 3I_Lisse_h2o, 3I_hanjie_h2o, 3I_combi_h2o, 3I_JUICE} are slightly shifted or following a different slope in the activity compared to the results in this work. These differences could be due to the reported hyperactivity of the comet \citep{3I_Biver, D_H_Nathan, 3I_JUICE} resulting in a contribution from an extended source of icy grains moving away from the nucleus. They would contribute to larger water production rates in the larger field of view of these instruments when compared to the small UVES slit. At the same time, the results are in very good agreement with the water production reported from JWST (red stars; \cite{JWST2025CO2}, \cite{JWST2026}, \cite{3I_JWST_Nathan}) and the power law fits very well with the activity reported from TRAPPIST \citep{jehin2025b,jehin2025a} and with the activity trend observed from SPHEREx \citep{3I_Lisse_ApJ, 3I_Lisse_h2o}. Variation in production rates from different facilities can also be attributed to the independent methodologies used for the computations. 

CN (0-0) was the emission observed the longest for about 15 epochs over 7 months. While the power law with a heliocentric index of $n = 6.04 \pm 0.10$ is consistent for the rates observed pre- and post-perihelion; there is a slight shift in the activity levels between both. The pre- and post-perihelion production rates reported by other studies \citep{jehin2025a, Rahatgaonkar2025UVES, hoogendam3I_Ni, 3I_medler_hoogendam, hoogendam2026keck, Zhao_3I_postper, 3I_lazzarin_3I} are seen to be in good agreement with the trend observed in this work. It is to be noted that despite \cite{Zhao_3I_postper} is using a different set of scale-lengths, which can result in marginally different absolute values, the overall activity trend is highly comparable.

As seen from Figure \ref{prod_rate}, owing to the limited number of pre-perihelion C$_2$ measurements and their relatively large uncertainties during the initial observations, the activity is not well understood, but is decreasing steeply post-perihelion. The emission followed a power law with a heliocentric index $n = 7.01 \pm 0.32$, which indicates a steeper evolution compared to CN. The pre-perihelion \citep{Rahatgaonkar2025UVES} and post-perihelion production rates reported by \cite{hoogendam2026keck} and \cite{jehin2025a} are consistent with the trend observed in this work. The C$_2$ production rates reported by \cite{Zhao_3I_postper} just after perihelion are in good agreement with those derived from UVES, although their last few measurements are slightly higher than our values. The origin of this difference is not clear and may reflect differences in the observational data, analysis methodology, or other assumptions adopted in the production-rate derivation. The larger slit used in their observations may also contribute, particularly given the hyperactive nature of 3I/ATLAS, as it samples a larger fraction of the extended coma and may therefore capture more of the extended C$_2$ emission (see Section \ref{abundance} for further details).

The C$_3$ (0-0) band was detected only on a very few epochs close to the perihelion. C$_3$ was observed to have the steepest activity rise with a heliocentric index $n = 9.43 \pm 2.83$. While the production rate reported by \cite{hoogendam2026keck} is in close agreement, the rates reported in \cite{Zhao_3I_postper} and \cite{jehin2025a} differ substantially. This could be because of the contribution from the strong NiI and FeI lines present near the C$_3$ emission band, especially in the post-perihelion epochs (see bottom panel in Figure \ref{panel2}). Using the metallic lines identified by \cite{3I_Damien_post}, we computed that the flux contribution from the metallic lines present in the C$_3$ region is twice the flux from C$_3$ itself. The  very low resolution used in \cite{Zhao_3I_postper} would result in the blend of these strong atomic lines with the C$_3$ band resulting in a higher observed flux and hence a higher production rate. In the case of TRAPPIST \citep{jehin2025a}, these strong atomic lines lie within the  C$_3$ filter passband, resulting also in higher observed rates. \cite{hoogendam2026keck} is able to extract the flux corresponding to the C$_3$ lines with the help of higher resolution. With UVES, we are able to extract precisely the emissions from C$_3$.  

The CH radical was detected only during a limited number of epochs close to perihelion, similar to C$_3$. Owing to the scarcity of CH observations in comets, its heliocentric evolution remains comparatively less explored than that of other common daughter species. Most published CH production rates have been reported by \citet{cochran_CH} and \citet{cochran_30years}. In 3I/ATLAS, the CH production rates exhibit a heliocentric power-law index of $n = 10.98 \pm 2.59$ during the pre-perihelion phase and $n = 8.87 \pm 0.96$ after perihelion. Although this may suggest a modest asymmetry in the activity, the small number of detections does not allow a robust assessment of its significance. 
Post-perihelion CH production rates reported by \citet{Zhao_3I_postper} show a large scatter; however, after averaging measurements obtained at nearby epochs, the resulting trend is broadly consistent with our values.

\cite{3I_kawakita2026} reports the absence of NH$_2$ emission, which is also true in our dataset. The high resolution and sensitivity of UVES in the blue side facilitates the detection of NH emissions between a forest of strong NiI lines. As for C$_3$, the presence of strong atomic lines within the bandpass of NH filter results in the higher production rates from TRAPPIST observations \citep{jehin2025a,jehin2025b}. NH was also detected only on a few epochs and followed an activity with heliocentric index $n = 5.87 \pm 1.22$.

\begin{figure}[h!]
  \centering
   \includegraphics[width=0.95\linewidth]{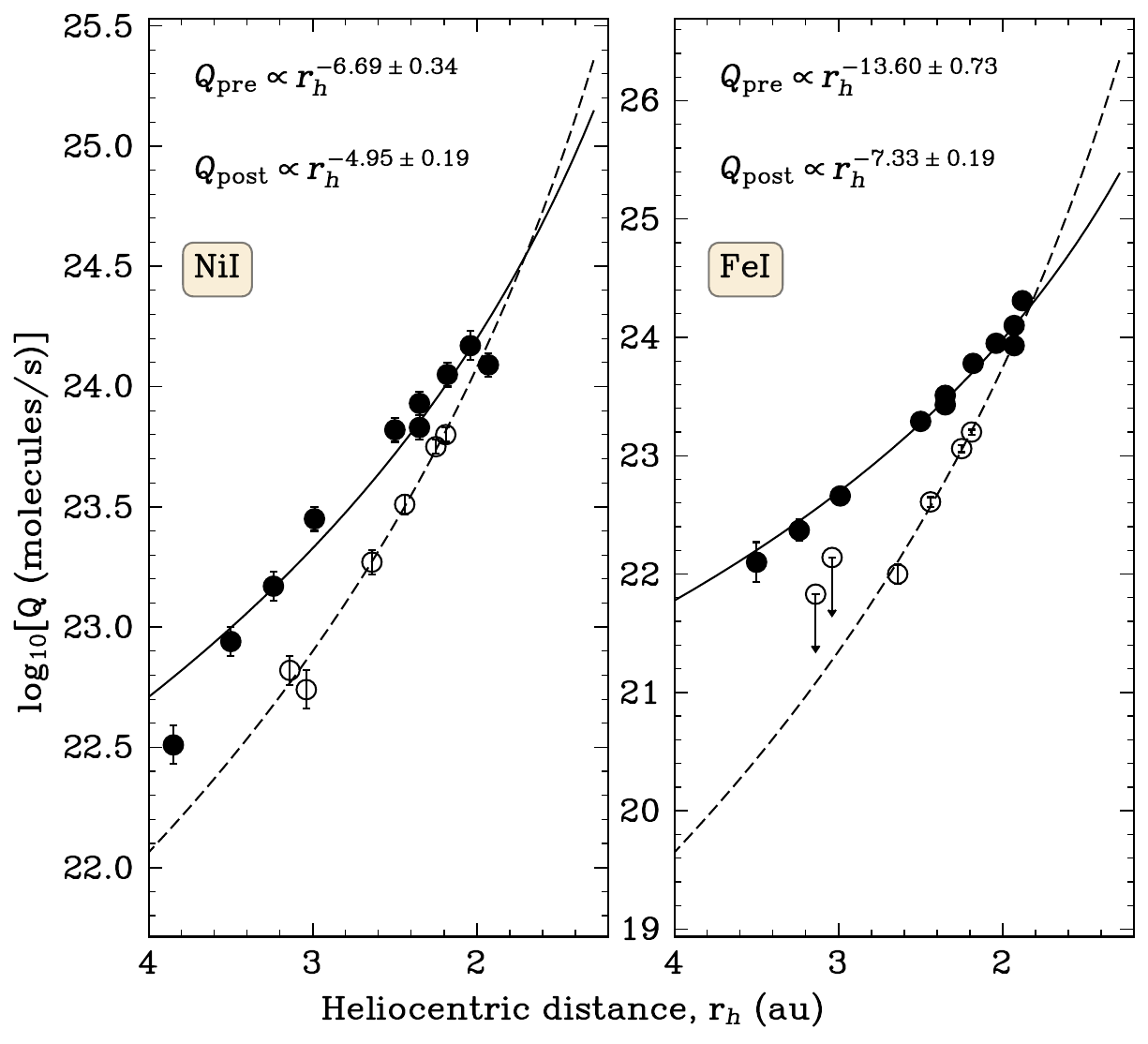}
      \caption{The power law fitting for NiI and FeI production rates of comet 3I/ATLAS reported in \cite{Hutsemekers2026NiFe} and \cite{3I_Damien_post}. Open symbols are for pre-perihelion measurements and filled for post-perihelion. The error bars correspond to 1$\sigma$ absolute errors.
              }
         \label{atoms_powerlaw}
   \end{figure}

The production rates of NiI and FeI computed from the same UVES dataset, and reported in \cite{Hutsemekers2026NiFe} and \cite{3I_Damien_post} for the pre- and post-perihelion epochs, were used to study the heliocentric dependence for these species. As seen in Figure \ref{atoms_powerlaw}, in contrast to the other observed species, the activity of both NiI and FeI emissions was highly asymmetric across perihelion. NiI/FeI had a steeper heliocentric power-law index of 6.69$\pm$0.34/13.60$\pm$0.73 during pre-perihelion epochs in comparison to an index 4.95$\pm$0.19/7.33$\pm$0.19 post-perihelion. Similar asymmetry with comparable slopes have been reported by \cite{Zhao_3I_postper}. 

The near symmetry of the power-law behaviour for the regular species argues for a relatively stable global response of the gaseous coma to solar heating with no seasonal effects, while the clear asymmetry of the metallic atoms, points to different release processes.

\subsection{Abundance ratios}\label{abundance}
\begin{figure*}[h!]
  \centering
   \includegraphics[width=0.93\linewidth]{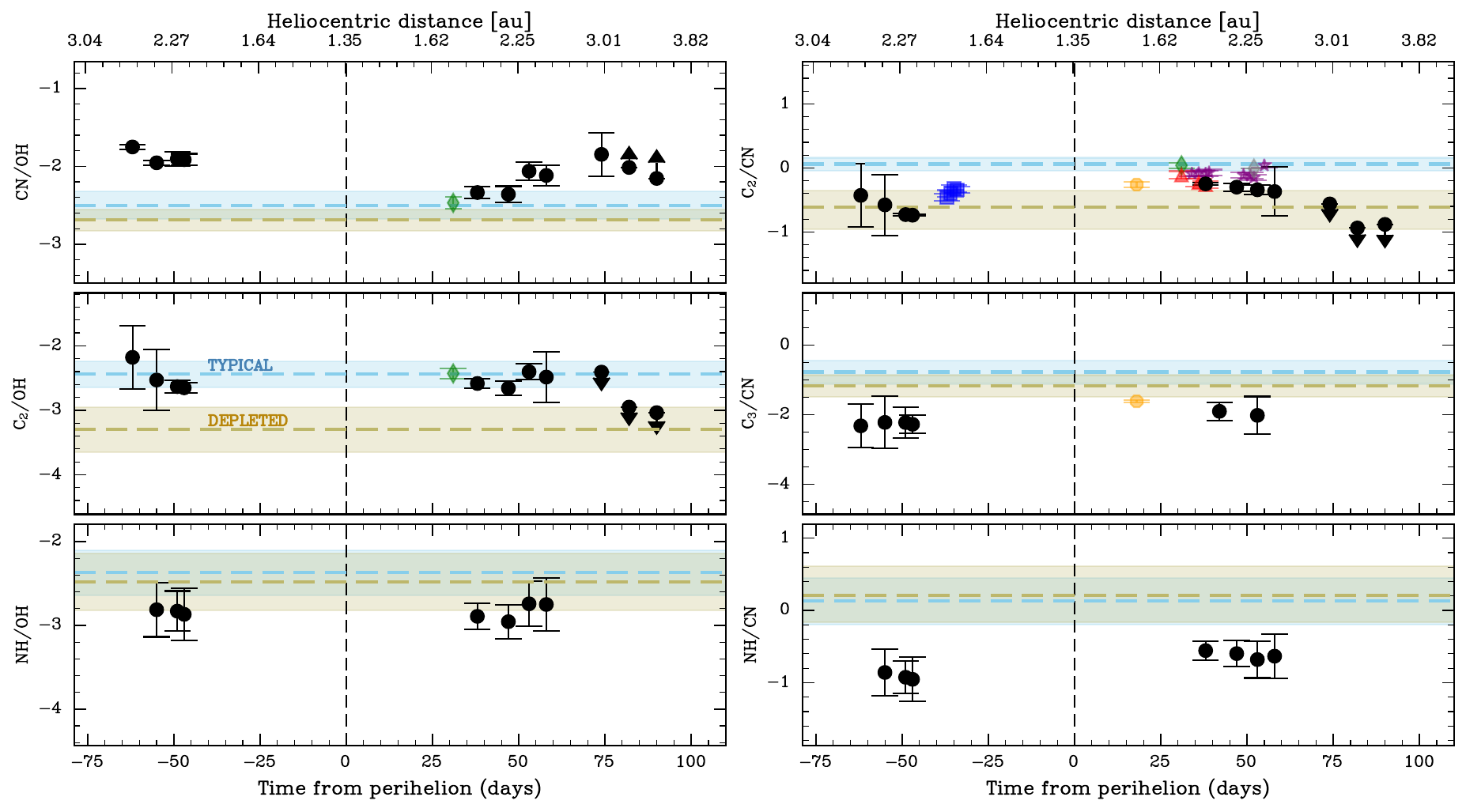}
      \caption{The logarithmic production rate ratios of comet 3I/ATLAS (black circle) as a function of days to perihelion and heliocentric distance compared with the rate ratios from other works: \cite{Rahatgaonkar2025UVES} (blue square),  \cite{hoogendam2026keck} (orange octagon), \cite{3I_kawakita2026} (red triangle), \cite{Zhao_3I_postper} (purple star), \cite{jehin2025a} (green diamond) and Lowell Observatory (David Schleicher - Private communication; gray diamond). The horizontal dashed lines and the shaded regions in blue and beige represent the mean ratios and range of ratios for comets with typical and depleted carbon
composition as defined by \cite{Ahearn_85}. The error bars correspond to 1$\sigma$ absolute errors.
              }
         \label{Qrateratio}
   \end{figure*}

Production rate ratios with respect to certain species is the best diagnosis to interpret the chemical peculiarities and the behaviour of the coma composition along its orbit. The variation of production rate ratios of each species with respect to OH and CN was analysed as a function of days to perihelion, as shown in Figure \ref{Qrateratio}. It is seen that the CN/OH decreased as the comet approached perihelion and then increased post-perihelion. 

As shown in Section \ref{powerlaw_sec}, the steeper heliocentric dependence of water production compared to CN could contribute to this trend and may indicate that CN parents are not predominantly released by the icy grains responsible for the comet's hyperactivity. While \cite{3I_Biver} reports 3I/ATLAS to be depleted in HCN/H$_2$O at 1.37 au, the CN/OH ratio measured here is typical.  However its decreasing trend suggests that CN/OH may become depleted at smaller heliocentric distances. Considering that the water production rate adopted by \cite{3I_Biver} from \cite{3I_JUICE} was derived using a comparable field of view, one may conclude that the measured extreme HCN/H$_2$O depletion is not merely a direct effect of aperture-dependent sampling. Similarly, the higher post-perihelion HCN/H$_2$O reported by \cite{3I_JWST_Nathan} at 2.4 au using a smaller field of view agrees with the typical CN/OH ratio we measure at comparable distance. These comparisons suggest that the heliocentric evolution as an effect of the hyperactive coma may affect the relative abundance ratios. The comparison with C/2017 K2 under similar observing geometries further shows that 3I/ATLAS produced substantially less OH relative to CN at comparable heliocentric distances (see Section \ref{17K2_sec} for further details).

The NH/CN ratio did not show any large variations and it clearly classifies the comet as NH-depleted according to the definition given in \cite{Ahearn_85}. This is in agreement to depletion reported in \cite{3I_kawakita2026}. The computed C$_3$/CN ratio lies in the depleted region and does not show any significant heliocentric trend. The ratio reported in \cite{hoogendam2026keck}, matches closely with the post-perihelion trend, but with a slightly higher value. The ratios reported by \cite{Zhao_3I_postper}, \cite{jehin2025a} and \cite{3I_kawakita2026} have not been included here as they point to a higher abundance which is due to the contamination from the strong atomic lines present in that region as discussed earlier. There is no contamination in other comets reported in \cite{Manfroid_Ni_Fe_comets} because their metallic lines are very weak, whereas those in 3I/ATLAS are roughly 100 times stronger \citep{3I_Damien_post}.

The C$_2$/CN ratio in comet 3I/ATLAS exhibits a pronounced dependence on heliocentric distance, with strongly depleted values at $r_h \gtrsim 2$ au, a slightly higher value post-perihelion and a subsequent decline as it receded beyond $\sim$2.2 au. Similar behaviour has been reported in other Solar System comets. \citet{schulz1998} found that comet 46P/Wirtanen classified as carbon-chain depleted beyond $\sim$2 au evolved towards a highly typical compositions at smaller heliocentric distances. Similar variation in the ratio for comet 46P has also been observed in its 2018 apparition \citep{moulane_46P, aravind_46P}. More recently, the production rate ratios of comet C/2017 K2 (PanSTARRS) reported by \cite{Said_17K2} showed that the comet displayed substantial drop in C$_2$/CN around $r_h \sim 2.0$ au during its outbound journey. Comparable heliocentric-distance-dependent trends have also been reported for comet C/2020 V2 (ZTF) \citep{Goldy_V2}, indicating that such variations are not uncommon among cometary comae.

Consequently, the C$_2$/CN evolution observed in 3I/ATLAS is naturally interpreted as the result of changing photochemical conditions, volatile release rates, and more effectively the hyperactivity of the comet as pointed out in the previous discussion. Furthermore, \cite{fellerec_12P} has shown that the effect of scale-lengths used for the computation of the ratios can be prominent at larger heliocentric distance. While Galactic Cosmic Ray processing of the outermost surface layers \citep{Maggiolo_2026} may contribute to the observed behaviour, the available observations do not require intrinsic compositional heterogeneity to explain the measured abundance ratios. 

This interpretation is further supported by other post-perihelion observations which reported a higher C$_2$/CN ratio in comparison to the pre-perihelion epochs \citep{jehin2025b, jehin2025a, Zhao_3I_postper, 3I_kawakita2026, hoogendam2026keck}. However, observations obtained with larger field of view \citep{jehin2025b,jehin2025a} and low-resolution long-slit spectroscopy \citep{3I_kawakita2026, Zhao_3I_postper}, reported slightly higher C$_2$/CN ratios than our UVES measurements, whereas the small-aperture observations of \citet{hoogendam2026keck} yielded values comparable to ours. Such aperture dependence is naturally explained if C$_2$ is produced from a parent molecule present in the icy grains releasing H$_2$O, allowing larger apertures to sample a greater fraction of the extended C$_2$ production region. The apparent discrepancies between different datasets therefore provide additional evidence to the hyperactivity of the comet which has been reported in other works \citep{3I_Biver, D_H_Nathan, 3I_JUICE}. 

The contrasting behaviour of the abundance ratios involving CN, C$_2$, and OH provides an important indication of the coupled evolution of these species. OH abundance is initially relatively low with respect to CN, resulting in a comparatively high CN/OH ratio; however, the steep increase in OH with decreasing heliocentric distance drives CN/OH towards the depleted range. Conversely, C$_2$ also increases steeply over the same heliocentric range, causing the C$_2$/CN ratio to increase towards the typical range. Despite these opposing trends in CN/OH and C$_2$/CN, the C$_2$/OH ratio remains consistently within the typical range defined by \cite{Ahearn_85}. This behaviour suggests that the increase in C$_2$ closely followed that of OH, such that their relative abundance remained largely unchanged even while both species evolved substantially with the heliocentric distance. The persistence of a typical C$_2$/OH ratio therefore indicates that C$_2$ and OH were largely coupled in their emission. In particular, if the enhanced C$_2$ production is associated with the release of C$_2$ parent species from icy grains, as expected in a hyperactive comet, the similar heliocentric evolution of C$_2$ and OH may indicate that their sources were activated in a coupled manner.

\subsection{3I/ATLAS versus other comets}\label{comparison}
3I/ATLAS is only the second active interstellar comet of its class. Competitively, 2I/Borisov was observed only at a few epochs due to its faint activity \citep{Borisov_highres, kareta_borisov, fitzsimmons2019, lin_borisov,borisov_COrich,borisov_waterprod, borisov_cordiner_CO, borisov_aravind}. The comparison of the absolute composition and heliocentric dependence of activity with respect to Solar System comets can put 3I/ATLAS in larger context to better understand its nature. A sample of comets observed using the TRAPPIST telescopes and belonging to different dynamical class (Long Period Comets (LPC); Halley Type Comets (HTC) and Jupiter Family Comets (JFC)) with their activities/slope published for atleast more than 10 epochs are then chosen for direct comparison. 
\cite{cochran_30years} provides a larger sample of comets for which the slopes of CN production rates have been reported. 

\begin{table}[h!]
    \centering
    \caption{Power-law slopes for the heliocentric dependence (Q $\propto$ r$^{-n}$) of production rates in 3I/ATLAS and Solar System comets belonging to various dynamical classes.}
    \setlength{\tabcolsep}{3.5pt}
		\resizebox{\linewidth}{!}{%
    \begin{tabular}{lcccccl}
        \hline
        \hline
        Comet & OH & NH & CN & C$_3$ & C$_2$ & Reference\\
        \hline
       46P$^a$ & 4.61$\pm$0.11&4.05$\pm$0.11 &4.39$\pm$0.02 & 5.44$\pm$0.05 & 5.13$\pm$0.03 & 1 \\
        103P$^a$& 5.88$\pm$0.67 & 5.21$\pm$0.17  & 4.19$\pm$0.09 & 5.43$\pm$0.33 & 4.17$\pm$0.14 & 2\\
        88P$^a$& 6.72$\pm$0.77&5.60$\pm$0.58&3.92$\pm$0.14&4.24$\pm$0.41&5.05$\pm$0.19&3\\
        12P$^b$ & 2.85$\pm$0.14 & 3.90$\pm$0.11 & 3.10$\pm$0.01 & 3.32$\pm$0.03 & 3.21$\pm$0.01 & 4\\
        C/2013 R1$^{c,f}$& 3.23$\pm$0.33 & 3.55$\pm$0.28 & 2.60$\pm$0.17 & 2.78$\pm$0.40 & 3.56$\pm$0.16 & 5\\
        C/2012 F6$^{c,f}$ & 2.52$\pm$0.05 & 2.81$\pm$0.05  & 2.54$\pm$0.03 & 2.65$\pm$0.03 & 2.97$\pm$0.03 & 6\\
        C/2012 V2$^c$&3.81$\pm$0.69&7.66$\pm$2.98&4.07$\pm$0.35&4.27$\pm$0.36&5.16$\pm$0.47&3\\
        C/2012 X1$^{c}$&3.47$\pm$0.19&3.32$\pm$0.23&1.73$\pm$0.08&2.24$\pm$0.08&2.44$\pm$0.10&3\\
        C/2017 K2$^d$ & 4.15$\pm$0.11 & 6.37$\pm$0.92 & 0.69$\pm$0.11  & 0.71$\pm$0.07 & 0.76$\pm$0.06 & 7\\
        C/2015 G2$^d$& 1.56$\pm$0.18&2.29$\pm$0.23&2.74$\pm$0.32&1.02$\pm$0.70&1.82$\pm$0.35&3\\ 
        C/2013 US10$^{d,f}$&1.45$\pm$0.35&1.84$\pm$0.12&0.74$\pm$0.14&1.16$\pm$0.09&1.35$\pm$0.13&3\\
        C/2013 A1$^{d}$&0.31$\pm$0.22&2.10$\pm$0.55&0.38$\pm$0.09&0.65$\pm$0.30&0.79$\pm$0.20&3\\
        3I$^e$& 8.47$\pm$0.23  & 5.87$\pm$1.22 & 6.04$\pm$0.10 & 9.43$\pm$2.83 & 7.01$\pm$0.32 & 8\\
    \hline
    \hline
    \multicolumn{7}{l}{$^a$ Jupiter Family Comet; $^b$ Halley Type Comets}\\
    \multicolumn{7}{l}{$^c$ Long period Comet - Dynamically Old; $^d$ Long period Comet - Dynamically New}\\
    \multicolumn{7}{l}{$^e$ Interstellar comet; $^f$ Pre-perihelion production rates have been adopted for reference}\\
     \multicolumn{7}{p{1.3\linewidth}}{References: (1) \cite{moulane_46P} (2) \cite{elise_103P} (3) \cite{Cyrielle_thesis} (4) \cite{mathieu_12P} (5) \cite{opitom_C2013R1} (6) \cite{opitom_C2012F6} (7) \cite{Said_17K2} (8) This work}\\
    \end{tabular}}
    \label{powerlaw_numbers}
\end{table}

Table \ref{powerlaw_numbers} compares the power law index obtained for the different species observed in 3I/ATLAS and comets belonging to different dynamical class. While the heliocentric range for 3I/ATLAS is about the same to the range considered for C/2017 K2 (PanSTARRS), C/2009 P1 (Garrardd) and 12P/Pons–Brooks, it is larger than the other comets considered in the sample. It is clear that the slopes of 3I/ATLAS are steep, but with slight resemblance to JFCs. \cite{Ahearn_85} quotes a median power law index of 2.77 for all gas species from all dynamical class together. But, it can be seen from various studies that many of the JFCs \citep{103P_knight, Lin_103P} have a steep heliocentric dependence as also seen in Table \ref{powerlaw_numbers}. The strongly carbon depleted comets discussed in \cite{JFC_cdepletion_schleicher} are also seen to have steep slopes. This same behaviour is also visible for the depleted JFCs reported in \cite{COMBI_waterprod_slope}.

\begin{figure*}[h!]
  \centering
   \includegraphics[width=0.92\linewidth]{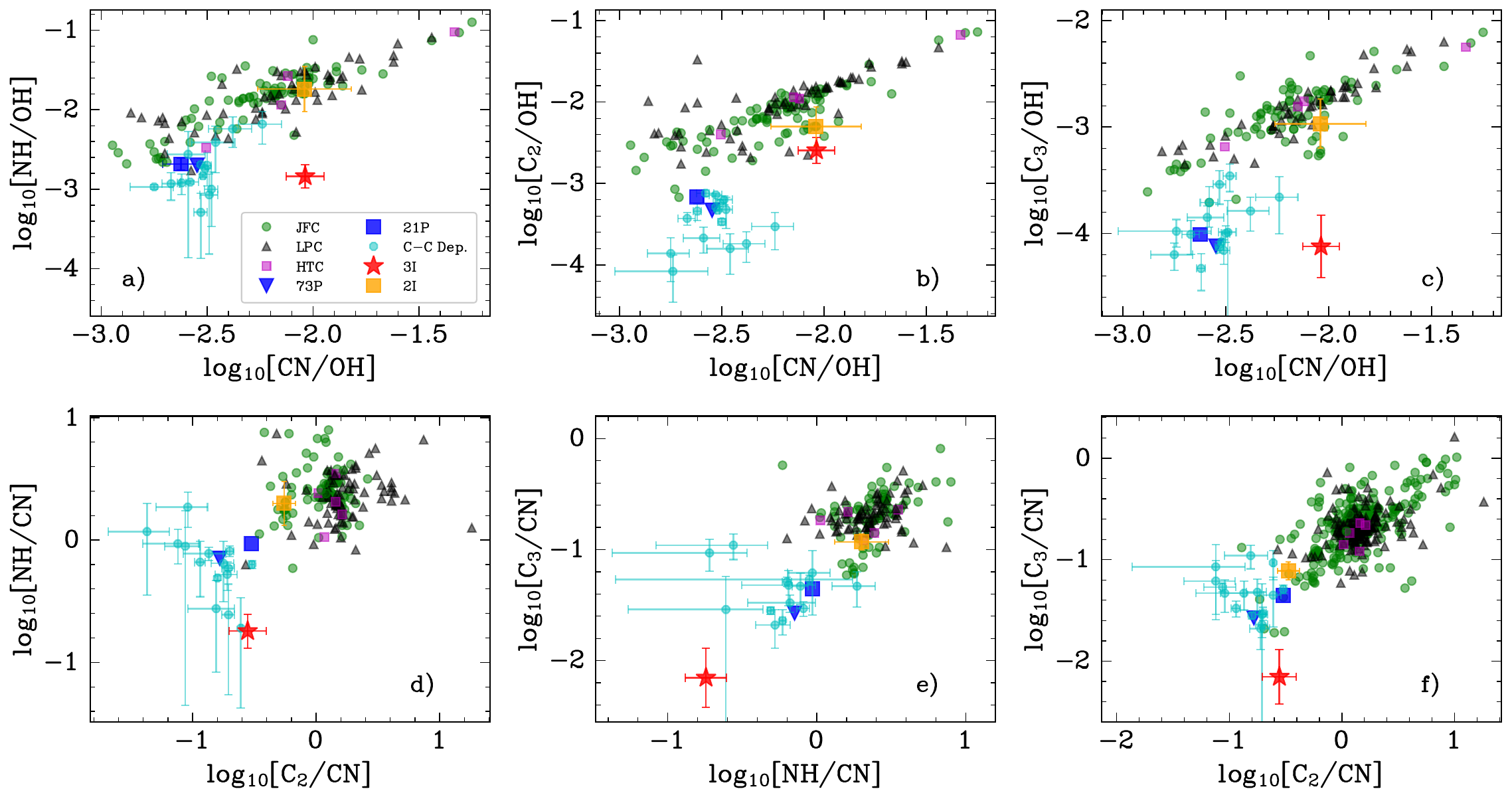}
      \caption{Mean logarithmic abundance ratios observed in 3I/ATLAS compared to a sample of strongly depleted comets \citep{21P_moulane, 21P_schleicher, 73P_schleicher, JFC_cdepletion_schleicher} and a large sample of Jupiter Family Comets (JFC), Long period Comets (LPC) and Halley Type Comets (HTC) \citep{cochran_30years, Aravind_PhDT, opitom_C2012F6, opitom_C2013R1, moulane_46P, Said_17K2, 21P_moulane, Manfroid_Ni_Fe_comets, 73P_schleicher, mathieu_12P, JFC_cdepletion_schleicher, langland-shula} along with the first interstellar comet 2I/Borisov \citep{Borisov_highres, Bair_2I}. The symbols are given in the first upper left panel. See Figure \ref{prod_rate_all_full} for the comparison of measured production rates.}
         \label{ratio_comp}
   \end{figure*}

In order to investigate whether the composition of 3I/ATLAS is distinct from Solar System comets, we first compared it with the dynamically new comet C/2017 K2 (see Section \ref{17K2_sec} for further details).  In order to further extend the investigation to compare the inter-dependence of production rates of the various molecular species, we compared production rates of 3I/ATLAS with those published for a large sample of Solar system comets. The sample includes 110 comets published in \cite{cochran_30years}, 26 comets published in \cite{langland-shula}, 17 strongly depleted comets reported in \cite{JFC_cdepletion_schleicher}, 21 comets studied as part of a thesis research \citep{Aravind_ASI_thesis, Aravind_PhDT}, 20 comets studied with UVES/VLT \citep{Manfroid_Ni_Fe_comets}, individual comets from TRAPPIST survey (46P/Wirtanen \citep{moulane_46P}; C/2013 R1 (Lovejoy) \citep{opitom_C2013R1}; C/2012 F6 (Lemmon) \citep{opitom_C2012F6}; 103P/Hartley 2 \citep{elise_103P}; C/2017 K2 (PanSTARRS) \citep{Said_17K2}; 21P/Giacobini–Zinner \citep{21P_moulane}; 12P/Pons–Brooks \citep{mathieu_12P}) and 73P/Schwassmann–Wachmann 3 \citep{73P_schleicher}. The production rates of the first active interstellar comet 2I/Borisov \citep{Borisov_highres}, reported from observations using UVES have also been included for a direct comparison.

The complete dataset with measured production rates in logarithmic scale is shown in Figure \ref{prod_rate_all_full}, whereas Figure \ref{stronglydepleted} represents a direct comparison of 3I/ATLAS with the strongly depleted Solar system comets \citep{JFC_cdepletion_schleicher} along with the Giacobini–Zinner (hereafter G-Z) type comets 21P/G-Z \citep{21P_schleicher, 21P_moulane} and 73P/Schwassmann–Wachmann \citep{73P_schleicher}. Figure \ref{ratio_comp} represents the comparison of the average logarithmic relative abundance with respect to OH and CN for 3I/ATLAS with the same for comets belonging to different classes in the sample.
As seen in Figure \ref{prod_rate_all_full}, most Solar System comets, irrespective of dynamical class, define tight correlations between the production rates of OH, NH, CN, C$_2$, and C$_3$, reflecting broadly similar volatile abundance ratios. The Jupiter Family Comet (JFC) distribution is broader in the case of C$_2$ vs CN (panel \textit{f} in Figure \ref{prod_rate_all_full}) as JFCs contain a much larger number of carbon-chain depleted comets in comparison to LPCs \citep{JFC_cdepletion_mathieu, JFC_cdepletion_schleicher}. Comet 12P, which underwent a strong outburst \citep{mathieu_12P} is also seen to tightly follow the correlations.

Although most comets define well-behaved linear correlations in the production-rate parameter space (Figure \ref{prod_rate_all_full}) or are clustered in a systematic manner in the logarithmic abundance ratio parameter space (Figure \ref{ratio_comp}), a distinct subgroup occupies a separate region in several of these diagrams. This population includes the prototype carbon-chain depleted comets 21P/G-Z \citep{21P_schleicher,21P_highres_depleted}, comet 73P/S-W 3, and the strongly depleted (C-C Dep.) JFCs identified by \citet{JFC_cdepletion_schleicher}. These objects are characterised by systematically reduced C$_2$ abundances and, pronounced NH depletion for the most depleted members. The close clustering of these comets in the abundance ratio parameter space suggests that they represent a common compositional class, broadly consistent with the G-Z type originally identified by \citet{Fink_comet_survey_2009}.

In several abundance ratio relations, 3I/ATLAS is shifted from the principal cometary cluster and lies close to the strongly depleted population as seen in Figure \ref{ratio_comp}. When the abundances are relative to OH, 3I/ATLAS lies below the main trends in the NH/OH and C$_3$/OH relations (Figure~\ref{ratio_comp}, panels \textit{a} and \textit{c}), approaching the strongly carbon-chain-depleted comets but with a higher CN relative to OH.

The distinction is also evident when the ratios are expressed relative to CN (panels \textit{d}–\textit{f}). 3I/ATLAS appears close to a few comets reported by \cite{JFC_cdepletion_schleicher}, with the largest NH and carbon depletion. The depletion observed in NH and C$_3$ appears to be stronger when CN is used as the reference species, with an average C$_2$/CN abundance comparable to that of the G-Z family comets (see panels \textit{d} and \textit{f} in Figure \ref{ratio_comp}).

Taken together, Figure \ref{ratio_comp} shows that 3I/ATLAS shares some characteristics of strongly carbon-chain-depleted or G-Z type Solar System comets, particularly its low NH and C$_3$ relative to OH and CN as also visible in Figure \ref{stronglydepleted}. However, its unusually high CN/OH ratio sets it apart from a straightforward carbon-chain-depleted or G-Z analogue, placing it in a distinct region in the taxonomy plot. The comparison with C/2017 K2 indicates that this significant variation in CN/OH ratio observed in 3I/ATLAS is primarily driven by the hyperactivity.

The limited measurements for 2I/Borisov remain within the broad distribution of carbon-depleted JFCs and show no significant shift in the relative abundance diagrams as seen in 3I/ATLAS. The distinct position of 3I/ATLAS also persists when the comparison is restricted to UVES measurements or to subsets analysed using common scale-lengths and g-factors, indicating that it is unlikely to arise from cross-survey systematics.

The resemblance with strongly carbon-depleted comets is particularly interesting. \citet{JFC_cdepletion_schleicher} argued that carbon-chain depletion reflects formation in particularly cold regions of the protosolar nebula, with an additional, lower-temperature threshold producing enhanced NH depletion. The similarity of the NH and C$_3$ depletion relative to OH with strongly depleted Solar System comets, together with its pronounced depletion in NH and C$_3$ relative to CN (as seen in panels \textit{e} and \textit{g} in Figure \ref{stronglydepleted}), may indicate formation under conditions colder than those experienced by the most depleted Solar System comets. This is  consistent with the idea that the chemical inventory of 3I/ATLAS was established in a very cold natal environment. This interpretation is independently supported by the unusually high $^{14}$N/$^{15}$N \citep{3I_Cyrielle_isotope, 3I_JWST_cordiner_2026} and water D/H \citep{D_H_Nathan} isotopic ratio reported for 3I/ATLAS, all of which point to a cold and distant formation region for the comet.

\section{Conclusions}
We present a homogeneous high-resolution spectroscopic monitoring of the interstellar comet 3I/ATLAS across perihelion, enabling a direct comparison of its gas activity before and after closest approach to the Sun. The dataset reveals a clear temporal evolution in the coma, with CN detected early, followed by the emergence of OH, NH, C$_3$, CH and C$_2$ as the comet brightened toward perihelion and then faded again afterward, which has always been the case for Solar System comets. The use of a single high-resolution instrument throughout the campaign provides a particularly robust framework for tracking the compositional changes without the systematic differences that often complicate multi-instrument comparisons.

We derive production rates for OH, NH, CN, C$_3$, CH and C$_2$ and find that all the species follow heliocentric-distance trends that are broadly consistent on both sides of the perihelion. This near-symmetry suggests that the main driver of the gas release remained similar across the apparition and point to a rather homogeneous surface of the nucleus. Unlike 2I which had both a seasonal effect causing a steep decline in water production rate post-perihelion \citep{borisov_waterprod} and fragmentation of the nucleus \citep{2I_split_jewitt}, 3I/ATLAS did not show any signs of these effects. The production rates of all the detected species display a noticeably stronger heliocentric dependence than is commonly observed in Solar System comets. This may reflect a combination of factors rather than a simple increase in sublimation efficiency. 

\cite{COMBI_waterprod_slope} discuss that the production-rate slopes are influenced by seasonal illumination, nucleus spin geometry, progressive penetration of the thermal wave, and surface evolution through dust-mantle removal and mass wasting, all of which modify the effective active area as the comet approaches the Sun. The heliocentric dependence of individual gaseous species reflects not only the incident solar flux but also the volatility and spatial distribution of their parent species. Progressive activation of new source regions, thermal-wave penetration into subsurface volatile reservoirs, and differences in sublimation behaviour between parent volatiles can all produce heliocentric power-law indices steeper than those expected from simple equilibrium sublimation \citep[e.g.,][]{fougere_2016, 67P_fornasier, jewitt_3I}. Such effects may have contributed to the relatively steep heliocentric dependences derived for several species in comet 3I/ATLAS in addition to the observed hyperactivity.

The detection of NH in 3I/ATLAS allows an optical constraint on the comet’s ammonia-related volatile composition and shows that NH and NH$_2$ are strongly depleted. We report an upper limit of the NH$_2$ production rate while no measurement of NH$_3$ has been reported in the infrared. NH$_2$ remained undetected in 3I/ATLAS in both the UVES and low-resolution spectra \citep{3I_kawakita2026}, whereas it was detected in both high- and low-resolution for 2I/Borisov \citep{Borisov_highres, Deam_2026}. This indicates that 3I/ATLAS exhibits a significantly lower NH$_2$/NH, suggesting differences in the nitrogen chemistry and/or photochemical evolution of the coma.

Our results fit within the rapidly growing literature on 3I/ATLAS. Overall, the UVES campaign shows that 3I/ATLAS is chemically dynamic, with a water signal consistent across facilities, a nitrogen-bearing component that is exceptionally weak, and a depleted carbon-chain ratio that evolves toward more typical values near perihelion before declining again. Even though the overall range of water production rates reported from different facilities are comparable, this analysis exposes the effects of differences between the methods and assumptions adopted for each techniques The observed heliocentric-distance dependence of the C$_2$/CN ratio in 3I/ATLAS, together with similar behaviour reported in several Solar System comets, strongly indicates that the measured variability is driven primarily by coma physics and rather than intrinsic nucleus composition. The consistency of this trend across independent datasets, combined with the aperture sensitivity of C$_2$ relative to CN, supports the presence of an extended source for C$_2$ (hyperactivity), where fragmentation of icy grains releases C$_2$-bearing parent molecules throughout the coma, whereas CN is produced predominantly from nucleus-originating parents.

The water production inferred from the forbidden red oxygen line is systematically higher than the OH-based estimate while the values from the two methods converge close to perihelion. This pattern suggests that, close to perihelion, water dominates the oxygen production, whereas at other epochs the oxygen lines likely contain a substantial contribution from CO$_2$-driven chemistry. The comparison with [OI]-based water production therefore provides a clear indication of the changing balance among H$_2$O, CO$_2$, and photochemical processes in the coma.

Finally, across multiple molecular production-rate and logarithmic relative abundance correlations, 3I/ATLAS consistently departs from the canonical Solar System comet cluster despite exhibiting absolute production rates comparable to those of Jupiter-family comets (see Figure \ref{prod_rate_all_full}). Its relative abundance relations place it closest to the population of strongly carbon-chain depleted comets, including the G-Z-type comets 21P/Giacobini–Zinner and 73P/Schwassmann–Wachmann 3. However, 3I/ATLAS exhibits a pronounced depletion of NH and C$_3$ relative to CN, closest to the most depleted comets reported by \cite{JFC_cdepletion_schleicher}, demonstrating that it is not simply an analogue of this population but an extension toward a more chemically extreme volatile inventory. Since the strongly depleted comets have been proposed to preserve primordial compositional signatures acquired in particularly cold regions of the protosolar nebula, the more pronounced imbalance observed in 3I/ATLAS further supports its formation in an exceptionally cold natal environment, consistent with recent isotopic studies.

Taken together, the high activity of 3I/ATLAS provided a rare opportunity to conduct the UVES observations to trace how an interstellar comet responds to solar heating across perihelion and to compare that response directly with well-studied Solar System comets. More broadly, this study offers an important observational benchmark for future space missions such as Comet Interceptor \citep{comet_interceptor}, which could characterise an interstellar comet in situ and will benefit from detailed remote-sensing constraints like those presented here.

\begin{acknowledgements}
      This publication uses data products from UVES, based on observations made with the ESO Very Large Telescope at the Paranal Observatory under programs 115.27ZL.001, 115.27ZJ.002 and 116.28NL.002. K. Aravind gratefully acknowledges support from the Wallonia-Brussels International (WBI) grant. D.H. and E.J. are Research Directors at the F.R.S.-FNRS. J.M. is honorary Research Director at the F.R.S-FNRS. We thank the staff at the Paranal observatory who made these observations possible in the service mode.
\end{acknowledgements}

\bibliographystyle{aa} 
\bibliography{reference.bib}

\begin{thebibliography}{130}
\expandafter\ifx\csname natexlab\endcsname\relax\def\natexlab#1{#1}\fi

\bibitem[{A'Hearn {et~al.}(1995)A'Hearn, Millis, Schleicher, Osip, \&
  Birch}]{Ahearn_85}
A'Hearn, M.~F., Millis, R.~C., Schleicher, D.~O., Osip, D.~J., \& Birch, P.~V.
  1995, Icarus, 118, 223

\bibitem[{{Ahuja} {et~al.}(2025){Ahuja}, {Aravind}, {Ganesh}, {Hmiddouch},
  {Donckt}, {Jehin}, {Sahu}, \& {Sivarani}}]{Goldy_V2}
{Ahuja}, G., {Aravind}, K., {Ganesh}, S., {et~al.} 2025, \mnras, 543, 1178

\bibitem[{{Aravind}(2023)}]{Aravind_PhDT}
{Aravind}, K. 2023, PhD thesis, Physical Research Laboratory, India

\bibitem[{{Aravind}(2024)}]{Aravind_ASI_thesis}
{Aravind}, K. 2024, in 42nd meeting of the Astronomical Society of India (ASI),
  Vol.~42, O75

\bibitem[{{Aravind} {et~al.}(2021){Aravind}, {Ganesh}, {Venkataramani}, {Sahu},
  {Angchuk}, {Sivarani}, \& {Unni}}]{borisov_aravind}
{Aravind}, K., {Ganesh}, S., {Venkataramani}, K., {et~al.} 2021, \mnras, 502,
  3491

\bibitem[{{Aravind} {et~al.}(2025){Aravind}, {Jehin}, {Hmiddouch}, {Vander
  Donckt}, {Ganesh}, {Rousselot}, {Hardy}, {Sahu}, {Manfroid}, \&
  {Benkhaldoun}}]{aravind_2020F3}
{Aravind}, K., {Jehin}, E., {Hmiddouch}, S., {et~al.} 2025, \aap, 701, A161

\bibitem[{{Aravind} {et~al.}(2024){Aravind}, {Venkataramani}, {Ganesh},
  {Jehin}, \& {Moulane}}]{aravind_46P}
{Aravind}, K., {Venkataramani}, K., {Ganesh}, S., {Jehin}, E., \& {Moulane}, Y.
  2024, Journal of Astrophysics and Astronomy, 45, 11

\bibitem[{{Bair} \& {Schleicher}(2025)}]{JFC_cdepletion_schleicher}
{Bair}, A.~N. \& {Schleicher}, D.~G. 2025, \psj, 6, 248

\bibitem[{{Bair} {et~al.}(2026){Bair}, {Schleicher}, {Kareta}, {Knight}, \&
  {Kelley}}]{Bair_2I}
{Bair}, A.~N., {Schleicher}, D.~G., {Kareta}, T., {Knight}, M.~M., \& {Kelley},
  M. S.~P. 2026, arXiv e-prints, arXiv:2608.23701

\bibitem[{{Ballester} {et~al.}(2000){Ballester}, {Modigliani}, {Boitquin},
  {Cristiani}, {Hanuschik}, {Kaufer}, \& {Wolf}}]{uves_pipeline}
{Ballester}, P., {Modigliani}, A., {Boitquin}, O., {et~al.} 2000, The
  Messenger, 101, 31

\bibitem[{Belyakov {et~al.}(2026)Belyakov, Wong, Bolin, Davis, Bromley, Lisse,
  \& Brown}]{JWST2026}
Belyakov, M., Wong, I., Bolin, B.~T., {et~al.} 2026, The Astrophysical Journal
  Letters, 1001, L11

\bibitem[{{Bhardwaj} \& {Raghuram}(2012)}]{Bhardwaj2012}
{Bhardwaj}, A. \& {Raghuram}, S. 2012, \apj, 748, 13

\bibitem[{{Biver} {et~al.}(2026){Biver}, {Bockel{\'e}e-Morvan}, {Moreno},
  {Crovisier}, {Paubert}, {Zakharov}, {Boissier}, {Cordiner}, \&
  {Roth}}]{3I_Biver}
{Biver}, N., {Bockel{\'e}e-Morvan}, D., {Moreno}, R., {et~al.} 2026, \aap, 708,
  L16

\bibitem[{{Bockel{\'e}e-Morvan} {et~al.}(2026){Bockel{\'e}e-Morvan}, {Poulet},
  {Langevin}, {Seignovert}, {Leyrat}, {Piccioni}, {Royer}, {Rodriguez},
  {d'Aversa}, {Brunetto}, {Carter}, {Cavali{\'e}}, {De Sanctis}, {Lellouch},
  {Migliorini}, {Pilorget}, {Quirico}, {Robert}, \& {Tosi}}]{3I_JUICE}
{Bockel{\'e}e-Morvan}, D., {Poulet}, F., {Langevin}, Y., {et~al.} 2026, \apjl,
  1006, L54

\bibitem[{{Bodewits} {et~al.}(2020){Bodewits}, {Noonan}, {Feldman},
  {Bannister}, {Farnocchia}, {Harris}, {Li}, {Mandt}, {Parker}, \&
  {Xing}}]{borisov_COrich}
{Bodewits}, D., {Noonan}, J.~W., {Feldman}, P.~D., {et~al.} 2020, Nature
  Astronomy, 4, 867

\bibitem[{Bolin {et~al.}(2025)Bolin, Belyakov, Fremling,
  {et~al.}}]{Bolin2025ATLAS}
Bolin, B.~T., Belyakov, M., Fremling, C., {et~al.} 2025, Monthly Notices of the
  Royal Astronomical Society: Letters, 542, L139–L144

\bibitem[{{Borisov} {et~al.}(2019){Borisov}, {Birtwhistle}, {Bacci},
  {Maestripieri}, {Chen}, {Green}, {Nakano}, {Sato}, \& {Durig}}]{Borisov2019}
{Borisov}, G., {Birtwhistle}, P., {Bacci}, P., {et~al.} 2019, Central Bureau
  Electronic Telegrams, 4666, 1

\bibitem[{{Cochran} \& {Barker}(1985)}]{cochran_CH}
{Cochran}, A.~L. \& {Barker}, E.~S. 1985, \icarus, 62, 72

\bibitem[{{Cochran} {et~al.}(2012){Cochran}, {Barker}, \&
  {Gray}}]{cochran_30years}
{Cochran}, A.~L., {Barker}, E.~S., \& {Gray}, C.~L. 2012, \icarus, 218, 144

\bibitem[{{Cochran} {et~al.}(2020){Cochran}, {Nelson}, \&
  {McKay}}]{21P_highres_depleted}
{Cochran}, A.~L., {Nelson}, T., \& {McKay}, A.~J. 2020, \psj, 1, 71

\bibitem[{{Cochran} \& {Schleicher}(1993)}]{cochran_OH_H2O}
{Cochran}, A.~L. \& {Schleicher}, D.~G. 1993, \icarus, 105, 235

\bibitem[{Combi {et~al.}(2019)Combi, Mäkinen, Bertaux, Quémerais, \&
  Ferron}]{COMBI_waterprod_slope}
Combi, M., Mäkinen, T., Bertaux, J.-L., Quémerais, E., \& Ferron, S. 2019,
  Icarus, 317, 610

\bibitem[{{Combi} {et~al.}(2026){Combi}, {M{\"a}kinen}, {Bertaux},
  {Qu{\'e}merais}, {Ferron}, {Lallement}, \& {Schmidt}}]{3I_combi_h2o}
{Combi}, M.~R., {M{\"a}kinen}, T., {Bertaux}, J.~L., {et~al.} 2026, \apjl, 998,
  L17

\bibitem[{Cordiner {et~al.}(2026)Cordiner, Roth, Micheli, Villanueva,
  Farnocchia, Charnley, Biver, Bockel{\'e}e-Morvan, Bodewits, Chandler,
  {et~al.}}]{3I_JWST_cordiner_2026}
Cordiner, M., Roth, N.~X., Micheli, M., {et~al.} 2026, Nature, 1

\bibitem[{{Cordiner} {et~al.}(2020){Cordiner}, {Milam}, {Biver},
  {Bockel{\'e}e-Morvan}, {Roth}, {Bergin}, {Jehin}, {Remijan}, {Charnley},
  {Mumma}, {Boissier}, {Crovisier}, {Paganini}, {Kuan}, \&
  {Lis}}]{borisov_cordiner_CO}
{Cordiner}, M.~A., {Milam}, S.~N., {Biver}, N., {et~al.} 2020, Nature
  Astronomy, 4, 861

\bibitem[{{Cordiner} {et~al.}(2025){Cordiner}, {Roth}, {Kelley}, {Bodewits},
  {Charnley}, {Drozdovskaya}, {Farnocchia}, {Micheli}, {Milam}, {Opitom},
  {Schwamb}, {Thomas}, \& {Bagnulo}}]{JWST2025CO2}
{Cordiner}, M.~A., {Roth}, N.~X., {Kelley}, M. S.~P., {et~al.} 2025, \apjl,
  991, L43

\bibitem[{{Coulson} {et~al.}(2026){Coulson}, {Kuan}, {Charnley}, {Cordiner},
  {Chuang}, {Lee}, {Lin}, {Milam}, {Pimpanuwat}, {Roth}, \&
  {{\.Z}{\'o}{\l}towski}}]{coulson_2026}
{Coulson}, I.~M., {Kuan}, Y.-J., {Charnley}, S.~B., {et~al.} 2026, \mnras, 546,
  stag063

\bibitem[{{Cremonese} {et~al.}(2020){Cremonese}, {Fulle}, {Cambianica},
  {Munaretto}, {Capria}, {La Forgia}, {Lazzarin}, {Migliorini}, {Boschin},
  {Milani}, {Aletti}, {Arlic}, {Bacci}, {Bacci}, {Bryssinck}, {Carosati},
  {Castellano}, {Buzzi}, {Di Rubbo}, {Facchini}, {Guido}, {Kugel}, {Ligustri},
  {Maestripieri}, {Mantero}, {Nicolas}, {Ochner}, {Perrella}, {Trabatti}, \&
  {Valvasori}}]{cremonese20202I}
{Cremonese}, G., {Fulle}, M., {Cambianica}, P., {et~al.} 2020, \apjl, 893, L12

\bibitem[{Crovisier {et~al.}(2025)Crovisier, Biver, \&
  Bockel{\'e}e-Morvan}]{Crovisier2025}
Crovisier, J., Biver, N., \& Bockel{\'e}e-Morvan, D. 2025, Comet 3I/ATLAS, cBET
  5625

\bibitem[{{de la Fuente Marcos} {et~al.}(2025){de la Fuente Marcos}, {Alarcon},
  {Licandro}, {Serra-Ricart}, {de Le{\'o}n}, {de la Fuente Marcos}, {Lombardi},
  {Tejero}, {Cabrera-Lavers}, {Guerra Arencibia}, \& {Ruiz Cejudo}}]{3I_10m}
{de la Fuente Marcos}, R., {Alarcon}, M.~R., {Licandro}, J., {et~al.} 2025,
  \aap, 700, L9

\bibitem[{Deam {et~al.}(2026)Deam, Bannister, Opitom, Knight, Ferellec,
  Ridden-Harper, Seligman, Fitzsimmons, Guilbert-Lepoutre, Jehin, Jorda,
  Marsset, Moulane, Rousselot, Vernazza, \& Yang}]{Deam_2026}
Deam, S.~E., Bannister, M.~T., Opitom, C., {et~al.} 2026, The Planetary Science
  Journal, 7, 88

\bibitem[{{Decock} {et~al.}(2013){Decock}, {Jehin}, {Hutsem{\'e}kers}, \&
  {Manfroid}}]{decock13}
{Decock}, A., {Jehin}, E., {Hutsem{\'e}kers}, D., \& {Manfroid}, J. 2013, 555,
  A34

\bibitem[{{Ferellec} {et~al.}(2024){Ferellec}, {Opitom}, {Donaldson}, {Fynbo},
  {Kokotanekova}, {Kelley}, \& {Lister}}]{fellerec_12P}
{Ferellec}, L., {Opitom}, C., {Donaldson}, A., {et~al.} 2024, \mnras, 534, 1816

\bibitem[{Ferellec {et~al.}(2026)Ferellec, Opitom, \& Snodgrass}]{Lea_3I}
Ferellec, L., Opitom, C., \& Snodgrass, C. 2026, Monthly Notices of the Royal
  Astronomical Society, stag1402

\bibitem[{{Festou} \& {Feldman}(1981)}]{festou81-oxygen}
{Festou}, M. \& {Feldman}, P.~D. 1981, 103, 154

\bibitem[{{Fink}(2009)}]{Fink_comet_survey_2009}
{Fink}, U. 2009, \icarus, 201, 311

\bibitem[{{Fink} \& {Hicks}(1996)}]{Fink1996}
{Fink}, U. \& {Hicks}, M.~D. 1996, \apj, 459, 729

\bibitem[{{Fitzsimmons} {et~al.}(2019){Fitzsimmons}, {Hainaut}, {Meech},
  {Jehin}, {Moulane}, {Opitom}, {Yang}, {Keane}, {Kleyna}, {Micheli}, \&
  {Snodgrass}}]{fitzsimmons2019}
{Fitzsimmons}, A., {Hainaut}, O., {Meech}, K.~J., {et~al.} 2019, \apjl, 885, L9

\bibitem[{{Fitzsimmons} {et~al.}(2024){Fitzsimmons}, {Meech}, {Matr{\`a}}, \&
  {Pfalzner}}]{comets_exo}
{Fitzsimmons}, A., {Meech}, K., {Matr{\`a}}, L., \& {Pfalzner}, S. 2024, in
  Comets III, ed. K.~J. {Meech}, M.~R. {Combi}, D.~{Bockel{\'e}e-Morvan}, S.~N.
  {Raymond}, \& M.~E. {Zolensky}, 731--766

\bibitem[{{Fitzsimmons} {et~al.}(2018){Fitzsimmons}, {Snodgrass}, {Rozitis},
  {Yang}, {Hyland}, {Seccull}, {Bannister}, {Fraser}, {Jedicke}, \&
  {Lacerda}}]{fitzsimmons2018}
{Fitzsimmons}, A., {Snodgrass}, C., {Rozitis}, B., {et~al.} 2018, Nature
  Astronomy, 2, 133

\bibitem[{{Fornasier} {et~al.}(2016){Fornasier}, {Mottola}, {Keller},
  {Barucci}, {Davidsson}, {Feller}, {Deshapriya}, {Sierks}, {Barbieri}, {Lamy},
  {Rodrigo}, {Koschny}, {Rickman}, {A'Hearn}, {Agarwal}, {Bertaux}, {Bertini},
  {Besse}, {Cremonese}, {Da Deppo}, {Debei}, {De Cecco}, {Deller}, {El-Maarry},
  {Fulle}, {Groussin}, {Gutierrez}, {G{\"u}ttler}, {Hofmann}, {Hviid}, {Ip},
  {Jorda}, {Knollenberg}, {Kovacs}, {Kramm}, {K{\"u}hrt}, {K{\"u}ppers},
  {Lara}, {Lazzarin}, {Moreno}, {Marzari}, {Massironi}, {Naletto}, {Oklay},
  {Pajola}, {Pommerol}, {Preusker}, {Scholten}, {Shi}, {Thomas}, {Toth},
  {Tubiana}, \& {Vincent}}]{67P_fornasier}
{Fornasier}, S., {Mottola}, S., {Keller}, H.~U., {et~al.} 2016, Science, 354,
  1566

\bibitem[{{Fougere} {et~al.}(2016){Fougere}, {Altwegg}, {Berthelier}, {Bieler},
  {Bockel{\'e}e-Morvan}, {Calmonte}, {Capaccioni}, {Combi}, {De Keyser},
  {Debout}, {Erard}, {Fiethe}, {Filacchione}, {Fink}, {Fuselier}, {Gombosi},
  {Hansen}, {H{\"a}ssig}, {Huang}, {Le Roy}, {Leyrat}, {Migliorini},
  {Piccioni}, {Rinaldi}, {Rubin}, {Shou}, {Tenishev}, {Toth}, \&
  {Tzou}}]{fougere_2016}
{Fougere}, N., {Altwegg}, K., {Berthelier}, J.-J., {et~al.} 2016, \mnras, 462,
  S156

\bibitem[{{Freudling} {et~al.}(2013){Freudling}, {Romaniello}, {Bramich},
  {Ballester}, {Forchi}, {Garc{\'{\i}}a-Dabl{\'o}}, {Moehler}, \&
  {Neeser}}]{esoreflex}
{Freudling}, W., {Romaniello}, M., {Bramich}, D.~M., {et~al.} 2013, \aap, 559,
  A96

\bibitem[{{Gray} {et~al.}(2025){Gray}, {Bagnulo}, {Borisov}, {Kwon}, {Cellino},
  {Kolokolova}, {Dorsey}, {Fedorets}, {Granvik}, {MacLennan}, {Mu{\~n}oz},
  {Bendjoya}, {Devog{\`e}le}, {Ieva}, {Penttil{\"a}}, \&
  {Muinonen}}]{Zuri2025Polarisation}
{Gray}, Z., {Bagnulo}, S., {Borisov}, G., {et~al.} 2025, \apjl, 992, L29

\bibitem[{{Guzik} \& {Drahus}(2021)}]{guzik2021Nature}
{Guzik}, P. \& {Drahus}, M. 2021, \nat, 593, 375

\bibitem[{{Guzik} {et~al.}(2020){Guzik}, {Drahus}, {Rusek}, {Waniak},
  {Cannizzaro}, \& {Pastor-Marazuela}}]{guzik_borisov}
{Guzik}, P., {Drahus}, M., {Rusek}, K., {et~al.} 2020, Nature Astronomy, 4, 53

\bibitem[{{Hardy} {et~al.}(2023){Hardy}, {Jehin}, {Rousselot},
  {Hutsem{\'e}kers}, \& {Manfroid}}]{Hardy_atlas}
{Hardy}, P., {Jehin}, E., {Rousselot}, P., {Hutsem{\'e}kers}, D., \&
  {Manfroid}, J. 2023, in LPI Contributions, Vol. 2851, LPI Contributions, 2148

\bibitem[{{Haser}(1957)}]{haser57}
{Haser}, L. 1957, Bull. Soc. R. Sci. Liege, 43, 740

\bibitem[{{Hemmen} {et~al.}(2026){Hemmen}, {Vander Donckt, M.}, {Jehin, E.},
  {Hmiddouch, S.}, {Aravind, K.}, {Manfroid, J.}, {Benkhaldoun, Z.}, {Jabiri,
  A.}, \& {Ganesh, S.}}]{elise_103P}
{Hemmen}, E., {Vander Donckt, M.}, {Jehin, E.}, {et~al.} 2026, A\&A, 710, A172

\bibitem[{{Hmiddouch} {et~al.}(2025){Hmiddouch}, {Jehin}, {Lippi}, {Vander
  Donckt}, {Aravind}, {Hutsem{\'e}kers}, {Manfroid}, {Jabiri}, {Moulane}, \&
  {Benkhaldoun}}]{Said_17K2}
{Hmiddouch}, S., {Jehin}, E., {Lippi}, M., {et~al.} 2025, \aap, 701, A61

\bibitem[{{Hoogendam} {et~al.}(2026){Hoogendam}, {Jones}, {Yang}, {Shappee},
  {Wray}, {Meech}, {Ashall}, {Desai}, {Hinkle}, {Hoffman}, {Medler}, {Pfeffer},
  \& {Zhao}}]{hoogendam2026keck}
{Hoogendam}, W.~B., {Jones}, D.~O., {Yang}, B., {et~al.} 2026, \apj, 1003, 245

\bibitem[{{Hoogendam} {et~al.}(2025){Hoogendam}, {Shappee}, {Wray}, {Yang},
  {Meech}, {Ashall}, {Desai}, {Hart}, {Hinkle}, {Hoffman}, {Hu}, {Jones},
  {Medler}, \& {Pfeffer}}]{hoogendam3I_Ni}
{Hoogendam}, W.~B., {Shappee}, B.~J., {Wray}, J.~J., {et~al.} 2025, arXiv
  e-prints, arXiv:2510.11779

\bibitem[{{Hui} {et~al.}(2020){Hui}, {Ye}, {F{\"o}hring}, {Hung}, \&
  {Tholen}}]{hui_borisov}
{Hui}, M.-T., {Ye}, Q.-Z., {F{\"o}hring}, D., {Hung}, D., \& {Tholen}, D.~J.
  2020, \aj, 160, 92

\bibitem[{{Hutsem{\'e}kers} {et~al.}(2026{\natexlab{a}}){Hutsem{\'e}kers},
  Manfroid, Jehin, Opitom, {et~al.}}]{Hutsemekers2026NiFe}
{Hutsem{\'e}kers}, D., Manfroid, J., Jehin, E., Opitom, C., {et~al.}
  2026{\natexlab{a}}, \aap, 676, A15

\bibitem[{{Hutsem{\'e}kers} {et~al.}(2026{\natexlab{b}}){Hutsem{\'e}kers},
  {Manfroid}, {Opitom}, {Jehin}, {Krishnakumar}, {Massa Fernandes},
  {Bannister}, {Bodewits}, {Dorsey}, {La Forgia}, \& {Murphy}}]{3I_Damien_post}
{Hutsem{\'e}kers}, D., {Manfroid}, J., {Opitom}, C., {et~al.}
  2026{\natexlab{b}}, arXiv e-prints, arXiv:2605.07652

\bibitem[{{Jehin} {et~al.}(2025{\natexlab{a}}){Jehin}, {Hardy}, {Blond Hanten},
  {Hutsem{\'e}kers}, {Manfroid}, \& {Sohy}}]{jehin_atlas_epsc}
{Jehin}, E., {Hardy}, P., {Blond Hanten}, E., {et~al.} 2025{\natexlab{a}}, in
  EPSC-DPS Joint Meeting 2025, Vol. 2025, EPSC--DPS2025--1777

\bibitem[{{Jehin} {et~al.}(2025{\natexlab{b}}){Jehin}, {Hmiddouch}, {Aravind},
  {Manfroid}, {Benkhaldoun}, \& {Jabiri}}]{jehin2025b}
{Jehin}, E., {Hmiddouch}, S., {Aravind}, K., {et~al.} 2025{\natexlab{b}}, The
  Astronomer's Telegram, 17538, 1

\bibitem[{{Jehin} {et~al.}(2025{\natexlab{c}}){Jehin}, {Hmiddouch}, {Aravind},
  {Manfroid}, {Benkhaldoun}, \& {Jabiri}}]{jehin2025a}
{Jehin}, E., {Hmiddouch}, S., {Aravind}, K., {et~al.} 2025{\natexlab{c}}, The
  Astronomer's Telegram, 17515, 1

\bibitem[{Jewitt {et~al.}(2019{\natexlab{a}})Jewitt, Agarwal, Hui, Li,
  Mutchler, \& Weaver}]{jewitt2019}
Jewitt, D., Agarwal, J., Hui, M.-T., {et~al.} 2019{\natexlab{a}}, The
  Astronomical Journal, 157, 65

\bibitem[{{Jewitt} {et~al.}(2017){Jewitt}, {Hui}, {Mutchler}, {Weaver}, {Li},
  \& {Agarwal}}]{Jewitt2017}
{Jewitt}, D., {Hui}, M.-T., {Mutchler}, M., {et~al.} 2017, \apjl, 847, L19

\bibitem[{{Jewitt} {et~al.}(2020){Jewitt}, {Kim}, {Mutchler}, {Weaver},
  {Agarwal}, \& {Hui}}]{2I_split_jewitt}
{Jewitt}, D., {Kim}, Y., {Mutchler}, M., {et~al.} 2020, \apjl, 896, L39

\bibitem[{Jewitt \& Luu(2025)}]{Jewitt2025ATLAS}
Jewitt, D. \& Luu, J. 2025, The Astrophysical Journal Letters, 950, L18

\bibitem[{{Jewitt} \& {Luu}(2025)}]{jewitt_3I}
{Jewitt}, D. \& {Luu}, J. 2025, \apjl, 994, L3

\bibitem[{Jewitt {et~al.}(2019{\natexlab{b}})Jewitt, Luu, Bolin,
  {et~al.}}]{Jewitt2019BorisovNature}
Jewitt, D., Luu, J., Bolin, B., {et~al.} 2019{\natexlab{b}}, Nature Astronomy,
  3, 924–930

\bibitem[{{Jewitt} \& {Seligman}(2023)}]{interstellar_interloppers}
{Jewitt}, D. \& {Seligman}, D.~Z. 2023, \araa, 61, 197

\bibitem[{{Jones} {et~al.}(2024){Jones}, {Snodgrass}, {Tubiana}, {K{\"u}ppers},
  {Kawakita}, {Lara}, {Agarwal}, {Andr{\'e}}, {Attree}, {Auster}, {Bagnulo},
  {Bannister}, {Beth}, {Bowles}, {Coates}, {Colangeli}, {Corral van Damme}, {Da
  Deppo}, {De Keyser}, {Della Corte}, {Edberg}, {El-Maarry}, {Faggi}, {Fulle},
  {Funase}, {Galand}, {Goetz}, {Groussin}, {Guilbert-Lepoutre}, {Henri},
  {Kasahara}, {Kereszturi}, {Kidger}, {Knight}, {Kokotanekova}, {Kolmasova},
  {Kossacki}, {K{\"u}hrt}, {Kwon}, {La Forgia}, {Levasseur-Regourd}, {Lippi},
  {Longobardo}, {Marschall}, {Morawski}, {Mu{\~n}oz}, {N{\"a}sil{\"a}},
  {Nilsson}, {Opitom}, {Pajusalu}, {Pommerol}, {Prech}, {Rando}, {Ratti},
  {Rothkaehl}, {Rotundi}, {Rubin}, {Sakatani}, {S{\'a}nchez}, {Simon Wedlund},
  {Stankov}, {Thomas}, {Toth}, {Villanueva}, {Vincent}, {Volwerk}, {Wurz},
  {Wielders}, {Yoshioka}, {Aleksiejuk}, {Alvarez}, {Amoros}, {Aslam},
  {Atamaniuk}, {Baran}, {Barci{\'n}ski}, {Beck}, {Behnke}, {Berglund},
  {Bertini}, {Bieda}, {Binczyk}, {Busch}, {Cacovean}, {Capria}, {Carr}, {Castro
  Mar{\'\i}n}, {Ceriotti}, {Chioetto}, {Chuchra-Konrad}, {Cocola}, {Colin},
  {Crews}, {Cripps}, {Cupido}, {Dassatti}, {Davidsson}, {De Roche}, {Deca},
  {Del Togno}, {Dhooghe}, {Donaldson Hanna}, {Eriksson}, {Fedorov},
  {Fern{\'a}ndez-Valenzuela}, {Ferretti}, {Floriot}, {Frassetto},
  {Fredriksson}, {Garnier}, {Gawe{\l}}, {G{\'e}not}, {Gerber}, {Glassmeier},
  {Granvik}, {Grison}, {Gunell}, {Hachemi}, {Hagen}, {Hajra}, {Harada},
  {Hasiba}, {Haslebacher}, {Herranz De La Revilla}, {Hestroffer}, {Hewagama},
  {Holt}, {Hviid}, {Iakubivskyi}, {Inno}, {Irwin}, {Ivanovski}, {Jansky},
  {Jernej}, {Jeszenszky}, {Jimen{\'e}z}, {Jorda}, {Kama}, {Kameda}, {Kelley},
  {Klepacki}, {Kohout}, {Kojima}, {Kowalski}, {Kuwabara}, {Ladno}, {Laky},
  {Lammer}, {Lan}, {Lavraud}, {Lazzarin}, {Le Duff}, {Lee}, {Lesniak}, {Lewis},
  {Lin}, {Lister}, {Lowry}, {Magnes}, {Markkanen}, {Martinez Navajas},
  {Martins}, {Matsuoka}, {Matyjasiak}, {Mazelle}, {Mazzotta Epifani}, {Meier},
  {Michaelis}, {Micheli}, {Migliorini}, {Millet}, {Moreno}, {Mottola},
  {Moutounaick}, {Muinonen}, {M{\"u}ller}, {Murakami}, {Murata}, {Myszka},
  {Nakajima}, {Nemeth}, {Nikolajev}, {Nordera}, {Ohlsson}, {Olesk}, {Ottacher},
  {Ozaki}, {Oziol}, {Patel}, {Savio Paul}, {Penttil{\"a}}, {Pernechele},
  {Peterson}, {Petraglio}, {Piccirillo}, {Plaschke}, {Polak}, {Postberg},
  {Proosa}, \& {Protopapa}}]{comet_interceptor}
{Jones}, G.~H., {Snodgrass}, C., {Tubiana}, C., {et~al.} 2024, \ssr, 220, 9

\bibitem[{{Kareta} {et~al.}(2020){Kareta}, {Andrews}, {Noonan}, {Harris},
  {Smith}, {O'Brien}, {Sharkey}, {Reddy}, {Springmann}, {Lejoly}, {Volk},
  {Conrad}, \& {Veillet}}]{kareta_borisov}
{Kareta}, T., {Andrews}, J., {Noonan}, J.~W., {et~al.} 2020, \apjl, 889, L38

\bibitem[{{Kawakita} {et~al.}(2026){Kawakita}, {Tsujimoto}, {Shinnaka},
  {Kobayashi}, {Ootsubo}, \& {Watanabe}}]{3I_kawakita2026}
{Kawakita}, H., {Tsujimoto}, K., {Shinnaka}, Y., {et~al.} 2026, \apjl, 1000,
  L60

\bibitem[{{Kawakita} \& {Watanabe}(2002)}]{kawakita_NH2_g}
{Kawakita}, H. \& {Watanabe}, J.-i. 2002, \apjl, 572, L177

\bibitem[{{Kawakita} {et~al.}(2001){Kawakita}, {Watanabe}, {Kinoshita}, {Abe},
  {Furusho}, {Izumiura}, {Yanagisawa}, \& {Masuda}}]{Kawakita_2001_NH2}
{Kawakita}, H., {Watanabe}, J.-I., {Kinoshita}, D., {et~al.} 2001, \pasj, 53,
  L5

\bibitem[{{Knight} \& {Schleicher}(2013)}]{103P_knight}
{Knight}, M.~M. \& {Schleicher}, D.~G. 2013, \icarus, 222, 691

\bibitem[{{Kurucz} {et~al.}(1984){Kurucz}, {Furenlid}, {Brault}, \&
  {Testerman}}]{kurucz_solar}
{Kurucz}, R.~L., {Furenlid}, I., {Brault}, J., \& {Testerman}, L. 1984, {Solar
  flux atlas from 296 to 1300 nm}

\bibitem[{{Langland-Shula} \& {Smith}(2011)}]{langland-shula}
{Langland-Shula}, L.~E. \& {Smith}, G.~H. 2011, \icarus, 213, 280

\bibitem[{{Lazzarin} {et~al.}(2026){Lazzarin}, {Mura}, {La Forgia},
  {Cremonese}, {Cambianica}, {Munaretto}, {Farina}, {Mazzotta Epifani}, {Ieva},
  \& {Dotto}}]{3I_lazzarin_3I}
{Lazzarin}, M., {Mura}, A.~C., {La Forgia}, F., {et~al.} 2026, \apjl, 998, L30

\bibitem[{{Li} {et~al.}(2026){Li}, {Shi}, {Hui}, \& {Shi}}]{3I_juncen_h2O}
{Li}, J., {Shi}, X., {Hui}, M.-T., \& {Shi}, J. 2026, \apjl, 999, L12

\bibitem[{{Lin} {et~al.}(2020){Lin}, {Lee}, {Gerdes}, {Adams}, {Becker},
  {Napier}, \& {Markwardt}}]{lin_borisov}
{Lin}, H.~W., {Lee}, C.-H., {Gerdes}, D.~W., {et~al.} 2020, \apjl, 889, L30

\bibitem[{{Lin} {et~al.}(2013){Lin}, {Lara}, \& {Ip}}]{Lin_103P}
{Lin}, Z.-Y., {Lara}, L.~M., \& {Ip}, W.-H. 2013, \aj, 146, 4

\bibitem[{{Lisse} {et~al.}(2025){Lisse}, {Bach}, {Bryan}, {Crill}, {Cukierman},
  {Dor{\'e}}, {Fabinsky}, {Faisst}, {Korngut}, {Melnick}, {Rustamkulov},
  {Tolls}, {Werner}, {Sitko}, {Champagne}, {Connelley}, {Emery}, {Fernandez},
  {Yang}, \& {the SPHEREx Science Team}}]{3I_lisse_2025Spherex}
{Lisse}, C.~M., {Bach}, Y.~P., {Bryan}, S., {et~al.} 2025, Research Notes of
  the American Astronomical Society, 9, 242

\bibitem[{{Lisse} {et~al.}(2026{\natexlab{a}}){Lisse}, {Bach}, {Bryan},
  {Korngut}, {Crill}, {Cukierman}, {Dor{\'e}}, {Cooray}, {Fabinsky}, {Faisst},
  {Hui}, {Melnick}, {Nguyen}, {Rustamkulov}, {Tolls}, {Wang}, {Werner}, \& {The
  Spherex Science Team}}]{3I_Lisse_h2o}
{Lisse}, C.~M., {Bach}, Y.~P., {Bryan}, S.~A., {et~al.} 2026{\natexlab{a}},
  Research Notes of the American Astronomical Society, 10, 26

\bibitem[{{Lisse} {et~al.}(2026{\natexlab{b}}){Lisse}, {Bach}, {Crill},
  {Korngut}, {Cukierman}, {Bryan}, {Cooray}, {Dowell}, {Werner}, {Hora},
  {Rustamkulov}, {Lee}, {Lim}, {Fernandez}, {Tolls}, {Reach}, {Dor{\'e}},
  {Zemcov}, {Bock}, {Cheng}, {Champagne}, {Choi}, {Connelley}, {Emery},
  {Everett}, {Faisst}, {Geem}, {Hui}, {Ishiguro}, {Jin}, {Jo}, {Mahlke},
  {Masters}, {Melnick}, {Nguyen}, {Paladini}, {Sitko}, \&
  {Yang}}]{3I_Lisse_ApJ}
{Lisse}, C.~M., {Bach}, Y.~P., {Crill}, B.~P., {et~al.} 2026{\natexlab{b}},
  \apjl, 1000, L52

\bibitem[{Maggiolo {et~al.}(2026)Maggiolo, Dhooghe, Gronoff, de~Keyser, \&
  Cessateur}]{Maggiolo_2026}
Maggiolo, R., Dhooghe, F., Gronoff, G.~P., de~Keyser, J., \& Cessateur, G.
  2026, The Astrophysical Journal Letters, 996, L34

\bibitem[{{Manfroid} {et~al.}(2021){Manfroid}, {Hutsem{\'e}kers}, \&
  {Jehin}}]{Manfroid_Ni_Fe_comets}
{Manfroid}, J., {Hutsem{\'e}kers}, D., \& {Jehin}, E. 2021, \nat, 593, 372

\bibitem[{{Manfroid} {et~al.}(2009){Manfroid}, {Jehin}, {Hutsem{\'e}kers},
  {Cochran}, {Zucconi}, {Arpigny}, {Schulz}, {St{\"u}we}, \&
  {Ilyin}}]{Manfroid2009}
{Manfroid}, J., {Jehin}, E., {Hutsem{\'e}kers}, D., {et~al.} 2009, \aap, 503,
  613

\bibitem[{{McKay} {et~al.}(2012){McKay}, {Chanover}, {Morgenthaler}, {Cochran},
  {Harris}, \& {Russo}}]{Mckay_2012_oxygen}
{McKay}, A.~J., {Chanover}, N.~J., {Morgenthaler}, J.~P., {et~al.} 2012,
  \icarus, 220, 277

\bibitem[{{McKay} {et~al.}(2020){McKay}, {Cochran}, {Dello Russo}, \&
  {DiSanti}}]{Mckay_borisov_oxygen}
{McKay}, A.~J., {Cochran}, A.~L., {Dello Russo}, N., \& {DiSanti}, M.~A. 2020,
  \apjl, 889, L10

\bibitem[{{Medler} {et~al.}(2026){Medler}, {Hoogendam}, {Ashall}, {Yang},
  {Wray}, {Shappee}, {Meech}, {Tucker}, {Auchettl}, {Desai}, {Hinkle},
  {Hoffman}, {Huber}, {Jones}, \& {Zhao}}]{3I_medler_hoogendam}
{Medler}, K., {Hoogendam}, W.~B., {Ashall}, C., {et~al.} 2026, \aj, 172, 25

\bibitem[{Meech {et~al.}(2017)Meech, Weryk, Micheli, Kleyna, Hainaut, Jedicke,
  Wainscoat, Chambers, Keane, Petric, Denneau, Magnier, Berger, Huber,
  Flewelling, Waters, Schunova-Lilly, \& Chastel}]{Meech2017}
Meech, K.~J., Weryk, R., Micheli, M., {et~al.} 2017, Nature, 552, 378

\bibitem[{{Meier} {et~al.}(1998){Meier}, {Wellnitz}, {Kim}, \&
  {A'Hearn}}]{NH_g}
{Meier}, R., {Wellnitz}, D., {Kim}, S.~J., \& {A'Hearn}, M.~F. 1998, \icarus,
  136, 268

\bibitem[{{Morgenthaler} {et~al.}(2007){Morgenthaler}, {Harris}, \&
  {Combi}}]{Morgenthaler_2007}
{Morgenthaler}, J.~P., {Harris}, W.~M., \& {Combi}, M.~R. 2007, \apj, 657, 1162

\bibitem[{{Morgenthaler} {et~al.}(2001){Morgenthaler}, {Harris}, {Scherb},
  {Anderson}, {Oliversen}, {Doane}, {Combi}, {Marconi}, \&
  {Smyth}}]{Morgenthaler_2001}
{Morgenthaler}, J.~P., {Harris}, W.~M., {Scherb}, F., {et~al.} 2001, \apj, 563,
  451

\bibitem[{{Moulane} {et~al.}(2023){Moulane}, {Jehin}, {Manfroid},
  {Hutsem{\'e}kers}, {Opitom}, {Shinnaka}, {Bodewits}, {Benkhaldoun}, {Jabiri},
  {Hmiddouch}, {Vander Donckt}, {Pozuelos}, \& {Yang}}]{moulane_46P}
{Moulane}, Y., {Jehin}, E., {Manfroid}, J., {et~al.} 2023, \aap, 670, A159

\bibitem[{{Moulane} {et~al.}(2020){Moulane}, {Jehin}, {Rousselot}, {Manfroid},
  {Shinnaka}, {Pozuelos}, {Hutsem{\'e}kers}, {Opitom}, {Yang}, \&
  {Benkhaldoun}}]{21P_moulane}
{Moulane}, Y., {Jehin}, E., {Rousselot}, P., {et~al.} 2020, \aap, 640, A54

\bibitem[{{Mura} {et~al.}(2026){Mura}, {Kawakita}, {La Forgia}, {Cremonese},
  {Lazzarin}, {Cambianica}, {Farina}, {Kobayashi}, {Manzini}, {Munaretto},
  {Ochner}, {Shinnaka}, \& {Tsujimoto}}]{Mura_CN}
{Mura}, A.~C., {Kawakita}, H., {La Forgia}, F., {et~al.} 2026, \aap, 712, A9

\bibitem[{{Opitom}(2016)}]{Cyrielle_thesis}
{Opitom}, C. 2016, PhD thesis, University of Liege, Belgium

\bibitem[{{Opitom} {et~al.}(2019){Opitom}, {Fitzsimmons}, {Jehin}, {Moulane},
  {Hainaut}, {Meech}, {Yang}, {Snodgrass}, {Micheli}, {Keane}, {Benkhaldoun},
  \& {Kleyna}}]{opitom_borisov}
{Opitom}, C., {Fitzsimmons}, A., {Jehin}, E., {et~al.} 2019, \aap, 631, L8

\bibitem[{{Opitom} {et~al.}(2020){Opitom}, {Guilbert-Lepoutre}, {Besse},
  {Yang}, \& {Snodgrass}}]{67P_muse_oxygen}
{Opitom}, C., {Guilbert-Lepoutre}, A., {Besse}, S., {Yang}, B., \& {Snodgrass},
  C. 2020, \aap, 644, A143

\bibitem[{{Opitom} {et~al.}(2021){Opitom}, {Jehin}, {Hutsem{\'e}kers},
  {Shinnaka}, {Manfroid}, {Rousselot}, {Raghuram}, {Kawakita}, {Fitzsimmons},
  {Meech}, {Micheli}, {Snodgrass}, {Yang}, \& {Hainaut}}]{Borisov_highres}
{Opitom}, C., {Jehin}, E., {Hutsem{\'e}kers}, D., {et~al.} 2021, \aap, 650, L19

\bibitem[{{Opitom} {et~al.}(2015{\natexlab{a}}){Opitom}, {Jehin}, {Manfroid},
  {Hutsem{\'e}kers}, {Gillon}, \& {Magain}}]{opitom_C2012F6}
{Opitom}, C., {Jehin}, E., {Manfroid}, J., {et~al.} 2015{\natexlab{a}}, \aap,
  574, A38

\bibitem[{{Opitom} {et~al.}(2015{\natexlab{b}}){Opitom}, {Jehin}, {Manfroid},
  {Hutsem{\'e}kers}, {Gillon}, \& {Magain}}]{opitom_C2013R1}
{Opitom}, C., {Jehin}, E., {Manfroid}, J., {et~al.} 2015{\natexlab{b}}, \aap,
  584, A121

\bibitem[{{Opitom} {et~al.}(2026){Opitom}, {Manfroid}, {Hutsem{\'e}kers},
  {Jehin}, {Knight}, {Aravind}, {Ferellec}, {Bodewits}, {Guzm{\'a}n},
  {Cordiner}, {Dorsey}, {La Forgia}, {Lippi}, {Murphy}, {Snodgrass}, \&
  {Bannister}}]{3I_Cyrielle_isotope}
{Opitom}, C., {Manfroid}, J., {Hutsem{\'e}kers}, D., {et~al.} 2026, Nature
  Astronomy [\eprint[arXiv]{2603.07187}]

\bibitem[{Opitom {et~al.}(2025)Opitom, Snodgrass, Jehin,
  {et~al.}}]{Opitom2025ATLAS}
Opitom, C., Snodgrass, C., Jehin, E., {et~al.} 2025, Monthly Notices of the
  Royal Astronomical Society: Letters, 544, L31–L35

\bibitem[{{Raghuram} \& {Bhardwaj}(2014)}]{Raghuram2014}
{Raghuram}, S. \& {Bhardwaj}, A. 2014, \aap, 566, A134

\bibitem[{Rahatgaonkar {et~al.}(2025)Rahatgaonkar, Carvajal, Puzia,
  {et~al.}}]{Rahatgaonkar2025UVES}
Rahatgaonkar, R., Carvajal, J.~P., Puzia, T.~H., {et~al.} 2025, The
  Astrophysical Journal Letters, 952, L5

\bibitem[{{Roth} {et~al.}(2026{\natexlab{a}}){Roth}, {Cordiner},
  {Bockel{\'e}e-Morvan}, {Biver}, {Crovisier}, {Milam}, {Lellouch},
  {Santos-Sanz}, {Lis}, {Qi}, {Foster}, {Boissier}, {Furuya}, {Moreno},
  {Charnley}, {Remijan}, {Kuan}, \& {Hart}}]{3IRadio2025}
{Roth}, N.~X., {Cordiner}, M.~A., {Bockel{\'e}e-Morvan}, D., {et~al.}
  2026{\natexlab{a}}, \apjl, 999, L32

\bibitem[{{Roth} {et~al.}(2026{\natexlab{b}}){Roth}, {Cordiner}, {Milam},
  {Villanueva}, {Charnley}, {Biver}, {Bockel{\'e}e-Morvan}, {Bodewits},
  {Bromley}, {Crovisier}, {Drozdovskaya}, {Faggi}, {Farnocchia}, {Furuya},
  {Kelley}, {Micheli}, {Noonan}, {Opitom}, {Schwamb}, \&
  {Thomas}}]{3I_JWST_Nathan}
{Roth}, N.~X., {Cordiner}, M.~A., {Milam}, S.~N., {et~al.} 2026{\natexlab{b}},
  \apjl, 1005, L5

\bibitem[{Salazar~Manzano {et~al.}(2025)Salazar~Manzano, Lin, Taylor,
  {et~al.}}]{SalazarManzano2025CN}
Salazar~Manzano, L.~E., Lin, H.~W., Taylor, A.~G., {et~al.} 2025, The
  Astrophysical Journal Letters, 953, L10

\bibitem[{{Salazar Manzano} {et~al.}(2026){Salazar Manzano},
  {Paneque-Carre{\~n}o}, {Cordiner}, {Bergin}, {Lin}, {Lis}, {Gerdes},
  {Bergner}, {Biver}, {Bockel{\'e}e-Morvan}, {Bodewits}, {Charnley},
  {Crovisier}, {Farnocchia}, {Guzm{\'a}n}, {Milam}, {Noonan}, {Remijan},
  {Roth}, \& {Tobin}}]{D_H_Nathan}
{Salazar Manzano}, L.~E., {Paneque-Carre{\~n}o}, T., {Cordiner}, M.~A.,
  {et~al.} 2026, Nature Astronomy [\eprint[arXiv]{2603.07026}]

\bibitem[{Santana-Ros {et~al.}(2025)Santana-Ros, Ivanova, Mykhailova,
  {et~al.}}]{SantanaRos2025ATLAS}
Santana-Ros, T., Ivanova, O., Mykhailova, S., {et~al.} 2025, \aap, 670, A7

\bibitem[{{Schleicher}(2010)}]{schleicher_CN_2010}
{Schleicher}, D.~G. 2010, \aj, 140, 973

\bibitem[{{Schleicher}(2022)}]{21P_schleicher}
{Schleicher}, D.~G. 2022, \psj, 3, 143

\bibitem[{{Schleicher} \& {A'Hearn}(1988)}]{OH_g}
{Schleicher}, D.~G. \& {A'Hearn}, M.~F. 1988, \apj, 331, 1058

\bibitem[{{Schleicher} \& {Bair}(2011)}]{73P_schleicher}
{Schleicher}, D.~G. \& {Bair}, A.~N. 2011, \aj, 141, 177

\bibitem[{{Schleicher} {et~al.}(1998){Schleicher}, {Millis}, \&
  {Birch}}]{Schleicher1998}
{Schleicher}, D.~G., {Millis}, R.~L., \& {Birch}, P.~V. 1998, \icarus, 132, 397

\bibitem[{{Schultz} {et~al.}(1992){Schultz}, {Li}, {Scherb}, \&
  {Roesler}}]{schultz_1992_oxygen}
{Schultz}, D., {Li}, G.~S.~H., {Scherb}, F., \& {Roesler}, F.~L. 1992, \icarus,
  96, 190

\bibitem[{{Schulz} {et~al.}(1998){Schulz}, {Arpigny}, {Manfroid}, {Stuewe},
  {Tozzi}, {Cremonese}, {Rembor}, \& {Peschke}}]{schulz1998}
{Schulz}, R., {Arpigny}, C., {Manfroid}, J., {et~al.} 1998, \aap, 335, L46

\bibitem[{{Seligman} {et~al.}(2025){Seligman}, {Micheli}, {Farnocchia},
  {Denneau}, {Noonan}, {Hsieh}, {Santana-Ros}, {Tonry}, {Auchettl}, {Conversi},
  {Devog{\`e}le}, {Faggioli}, {Feinstein}, {Fenucci}, {Ferrais}, {Frincke},
  {Gillon}, {Hainaut}, {Hart}, {Hoffman}, {Holt}, {Hoogendam}, {Huber},
  {Jehin}, {Kareta}, {Keane}, {Kelley}, {Lister}, {Mandt}, {Manfroid},
  {Mar{\v{c}}eta}, {Meech}, {Amine Miftah}, {Morgan}, {Oca{\~n}a},
  {Pe{\~n}a-Asensio}, {Shappee}, {Siverd}, {Taylor}, {Tucker}, {Wainscoat},
  {Weryk}, {Wray}, {Yaginuma}, {Yang}, {Ye}, \& {Zhang}}]{seligman2025}
{Seligman}, D.~Z., {Micheli}, M., {Farnocchia}, D., {et~al.} 2025, \apjl, 989,
  L36

\bibitem[{{Shinnaka} {et~al.}(2026){Shinnaka}, {Tsujimoto}, {Kawakita},
  {Kobayashi}, {Watanabe}, \& {Ootsubo}}]{Shinnaka_3I_oxygen}
{Shinnaka}, Y., {Tsujimoto}, K., {Kawakita}, H., {et~al.} 2026, \aj, 171, 281

\bibitem[{{Tan} {et~al.}(2026){Tan}, {Yan}, \& {Li}}]{3I_hanjie_h2o}
{Tan}, H., {Yan}, X., \& {Li}, J.-Y. 2026, \apjl, 998, L22

\bibitem[{{van Dokkum}(2001)}]{LAcosmic}
{van Dokkum}, P.~G. 2001, \pasp, 113, 1420

\bibitem[{{van Dokkum} {et~al.}(2012){van Dokkum}, {Bloom}, \&
  {Tewes}}]{lacosmic2012}
{van Dokkum}, P.~G., {Bloom}, J., \& {Tewes}, M. 2012, {L.A.Cosmic: Laplacian
  Cosmic Ray Identification}, Astrophysics Source Code Library, record
  ascl:1207.005

\bibitem[{{Vander Donckt} {et~al.}(2023){Vander Donckt}, {Aravind}, {Jehin},
  {Ganesh}, {Hmiddouch}, {Moulane}, {Benkhaldoun}, {Jabiri}, {Sahu}, \&
  {Sivarani}}]{JFC_cdepletion_mathieu}
{Vander Donckt}, M., {Aravind}, K., {Jehin}, E., {et~al.} 2023, in LPI
  Contributions, Vol. 2851, Asteroids, Comets, Meteors Conference, 2461

\bibitem[{{Vander Donckt} {et~al.}(2026){Vander Donckt}, {Jehin}, {Aravind},
  {Adami}, {Hmiddouch}, {Manfroid}, {Ganesh}, {Benkhaldoun}, \&
  {Delsanti}}]{mathieu_12P}
{Vander Donckt}, M., {Jehin}, E., {Aravind}, K., {et~al.} 2026, \aap, 705, A89

\bibitem[{{Xing} {et~al.}(2020){Xing}, {Bodewits}, {Noonan}, \&
  {Bannister}}]{borisov_waterprod}
{Xing}, Z., {Bodewits}, D., {Noonan}, J., \& {Bannister}, M.~T. 2020, \apjl,
  893, L48

\bibitem[{{Xing} {et~al.}(2025){Xing}, {Oset}, {Noonan}, \&
  {Bodewits}}]{3I_Xing_h2O}
{Xing}, Z., {Oset}, S., {Noonan}, J., \& {Bodewits}, D. 2025, \apjl, 991, L50

\bibitem[{{Yang} {et~al.}(2020){Yang}, {Kelley}, {Meech}, {Keane}, {Protopapa},
  \& {Bus}}]{yang_waterice}
{Yang}, B., {Kelley}, M. S.~P., {Meech}, K.~J., {et~al.} 2020, \aap, 634, L6

\bibitem[{Yang {et~al.}(2025)Yang, Meech, Connelley, Zhao,
  {et~al.}}]{Yang2025NIR}
Yang, B., Meech, K.~J., Connelley, M., Zhao, R., {et~al.} 2025, The
  Astrophysical Journal Letters, 951, L12

\bibitem[{{Ye} {et~al.}(2017){Ye}, {Zhang}, {Kelley}, \&
  {Brown}}]{ye17-oumuamua}
{Ye}, Q.-Z., {Zhang}, Q., {Kelley}, M.~S.~P., \& {Brown}, P.~G. 2017, 851, L5

\bibitem[{Zhao {et~al.}(2026)Zhao, Zhang, Yang, Fan, Wang, Huang, \&
  Liu}]{Zhao_3I_postper}
Zhao, R., Zhang, X., Yang, B., {et~al.} 2026, The Astrophysical Journal
  Letters, 1004, L24

\bibitem[{{Zheltobryukhov} {et~al.}(2025){Zheltobryukhov}, {Zubko},
  {Tsuranova}, {Kochergin}, {Chornaya}, \& {Videen}}]{2025RNAAS}
{Zheltobryukhov}, M., {Zubko}, E., {Tsuranova}, M., {et~al.} 2025, Research
  Notes of the American Astronomical Society, 9, 338

\bibitem[{{Zubko} \& {Videen}(2025)}]{zubko2025polarisation}
{Zubko}, E. \& {Videen}, G. 2025, \mnras, 544, L156

\end{thebibliography}

\begin{appendix}
\onecolumn
\section{High-resolution plots of emissions detected in 3I/ATLAS}

\begin{figure*}[h!]
\begin{subfigure}{0.49\linewidth}
\centering
\includegraphics[width = 0.91\textwidth]{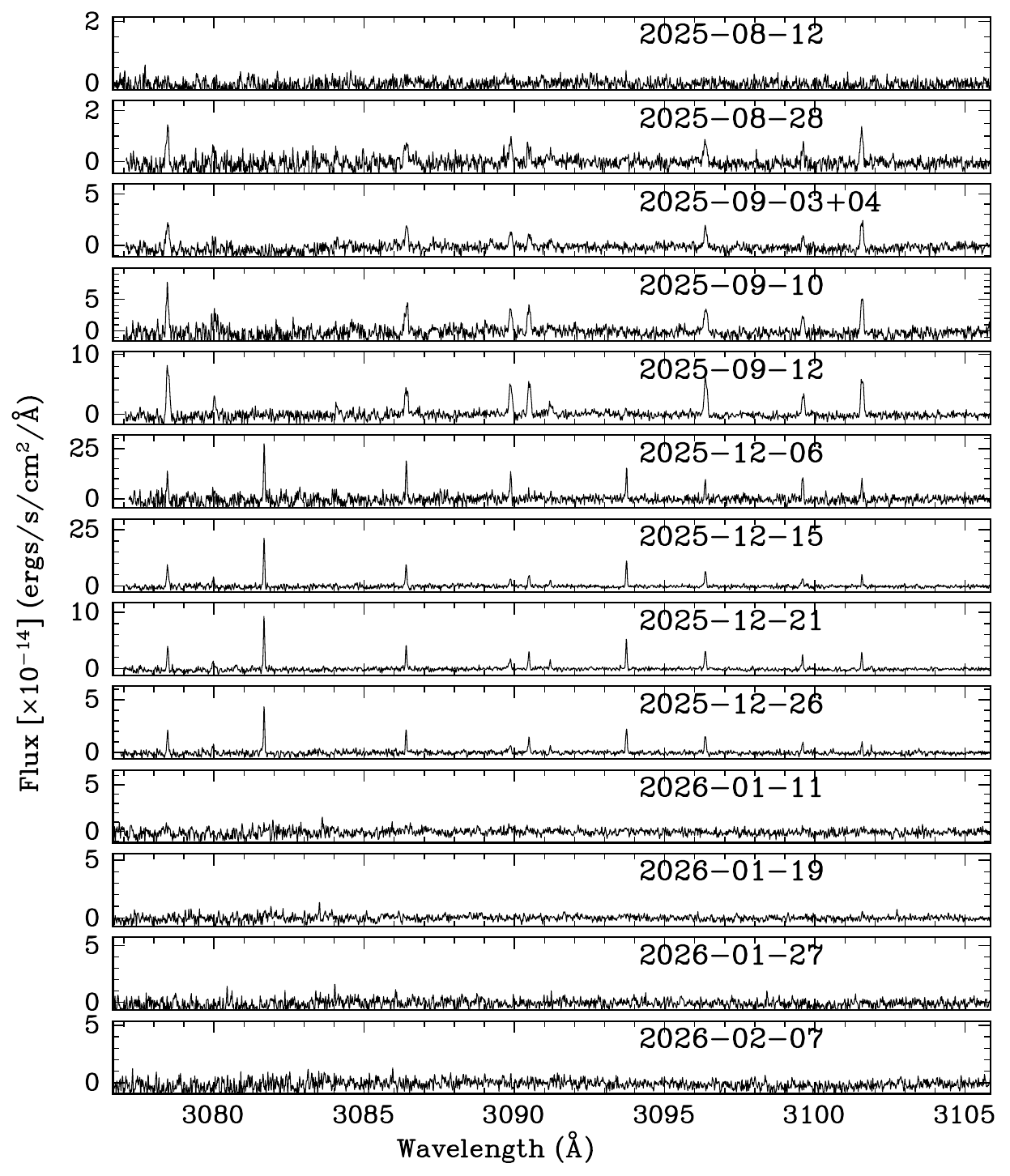}
\label{OH_panel}
\end{subfigure}\hfill
\begin{subfigure}{0.49\linewidth}
\centering
\includegraphics[width = 0.99\textwidth]{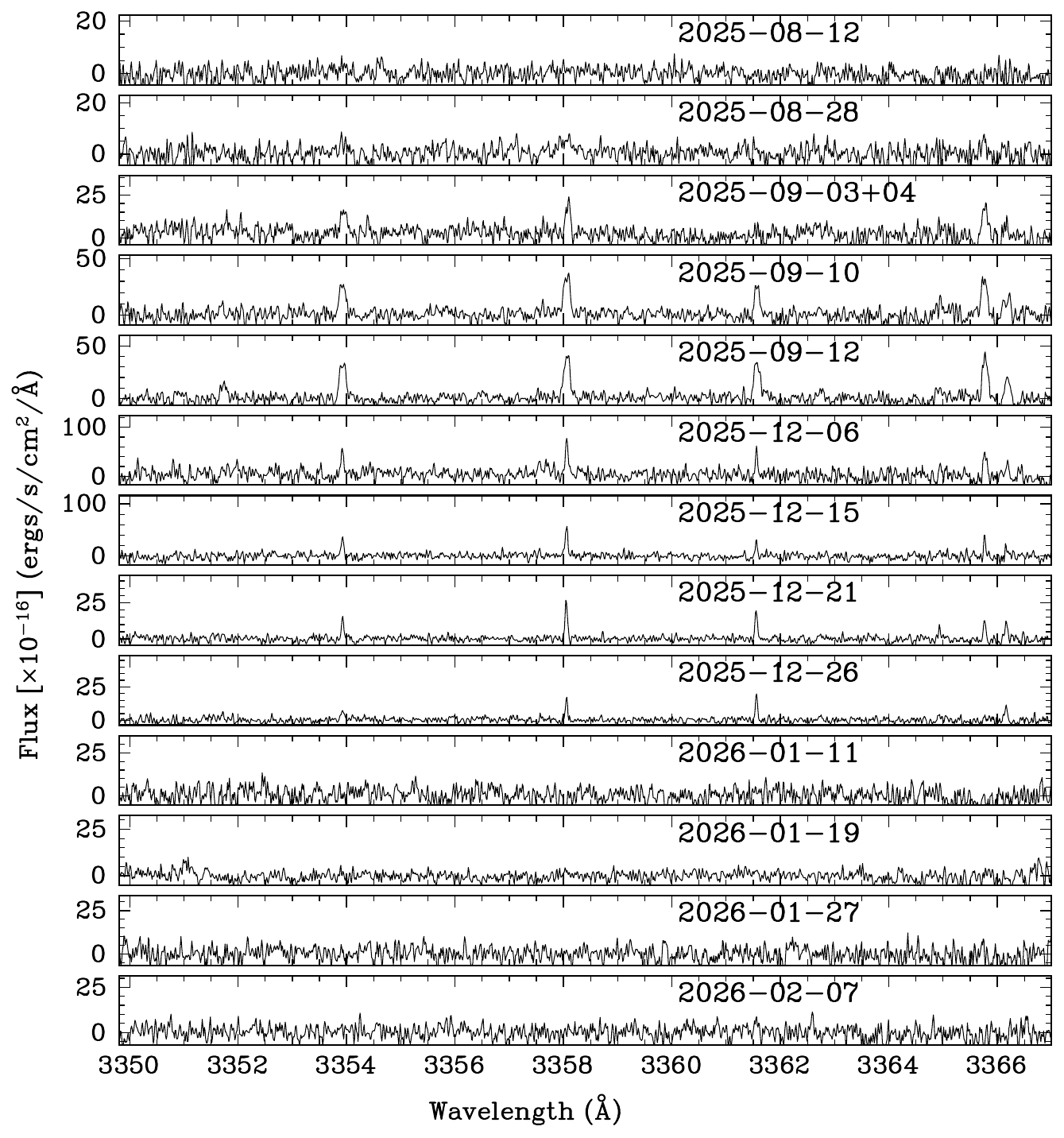}
\label{NH_panel}
\end{subfigure}\hfill
\begin{subfigure}{0.99\linewidth}
\centering
\includegraphics[width = 0.99\textwidth]{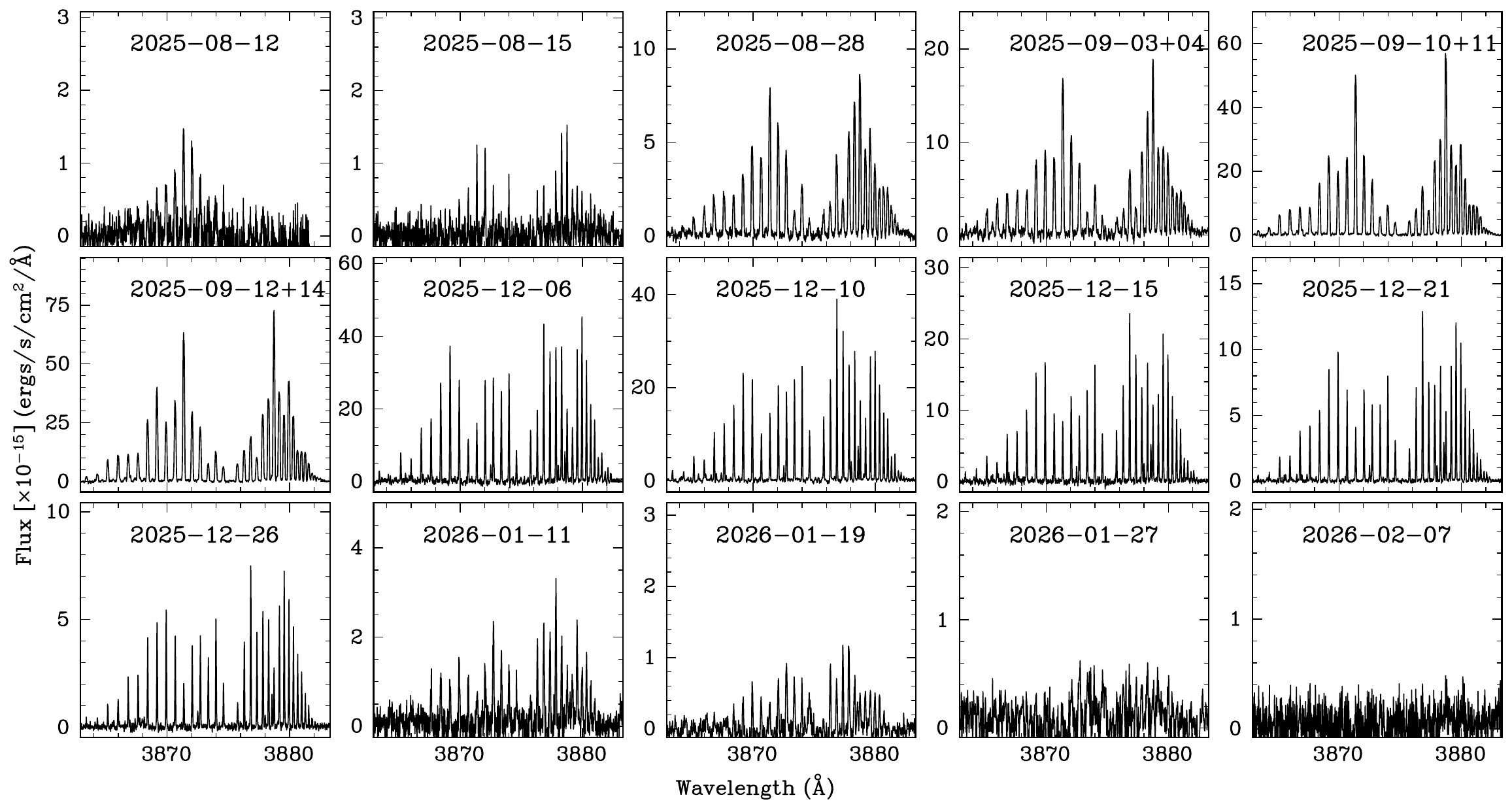}
\label{CN_panel}
\end{subfigure}\hfill
\caption{Evolution of the OH, NH, and CN emissions detected in 3I/ATLAS during the complete observing campaign. \textit{Left}: OH emissions detected in the UVES spectra on different epochs. \textit{Right}: NH emissions detected in the UVES spectra on different epochs. \textit{Bottom}: CN emissions detected in the UVES spectra on different epochs.}
\label{panel1}
\end{figure*}


\begin{figure*}
\begin{subfigure}{0.49\linewidth}
\centering
\includegraphics[width = 0.99\textwidth]{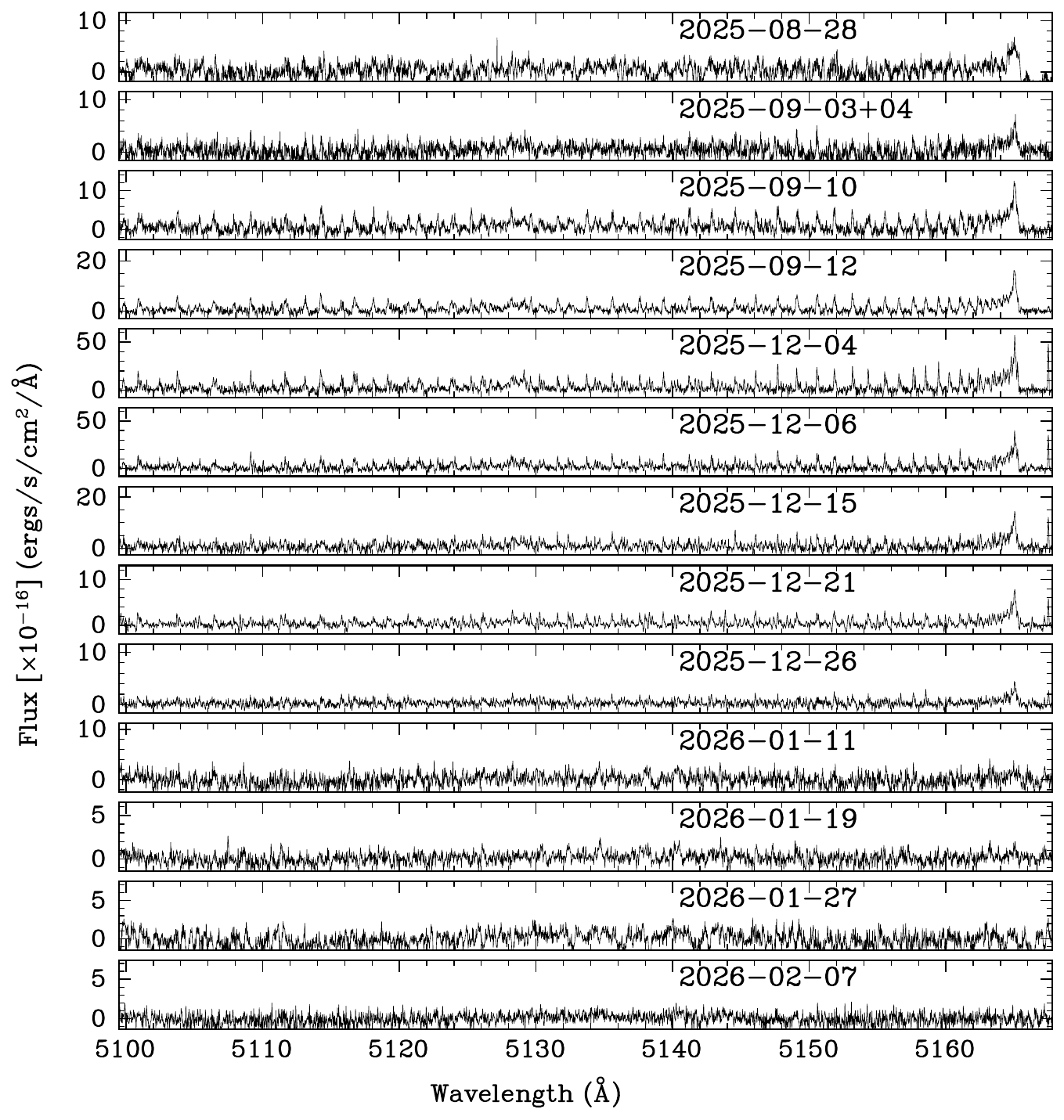}
\label{C2_panel}
\end{subfigure}\hfill
\begin{subfigure}{0.49\linewidth}
\centering
\includegraphics[width = 0.88\textwidth]{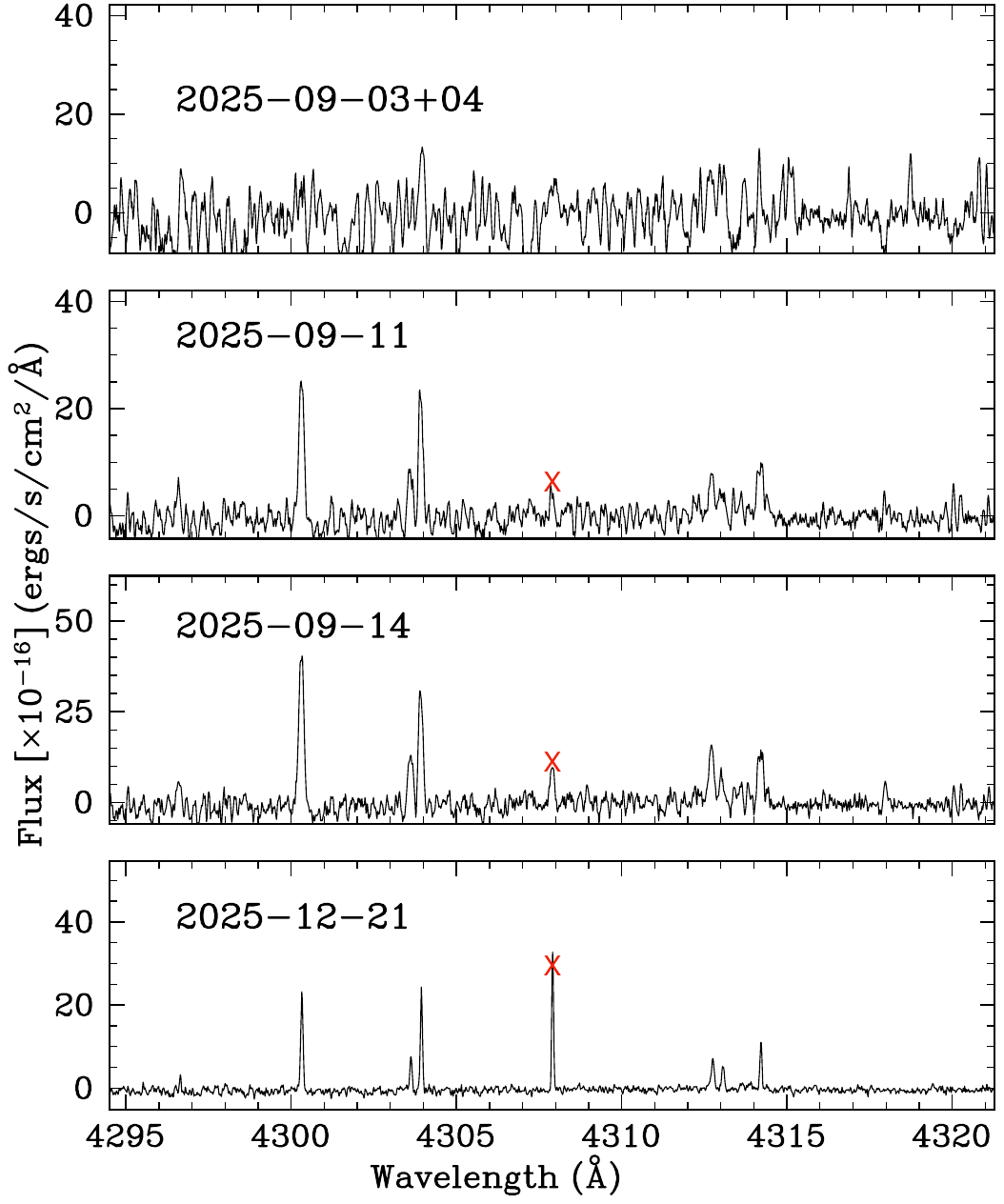}
\label{CH_panel}
\end{subfigure}\hfill
\begin{subfigure}{0.99\linewidth}
\centering
\includegraphics[width = 0.65\textwidth]{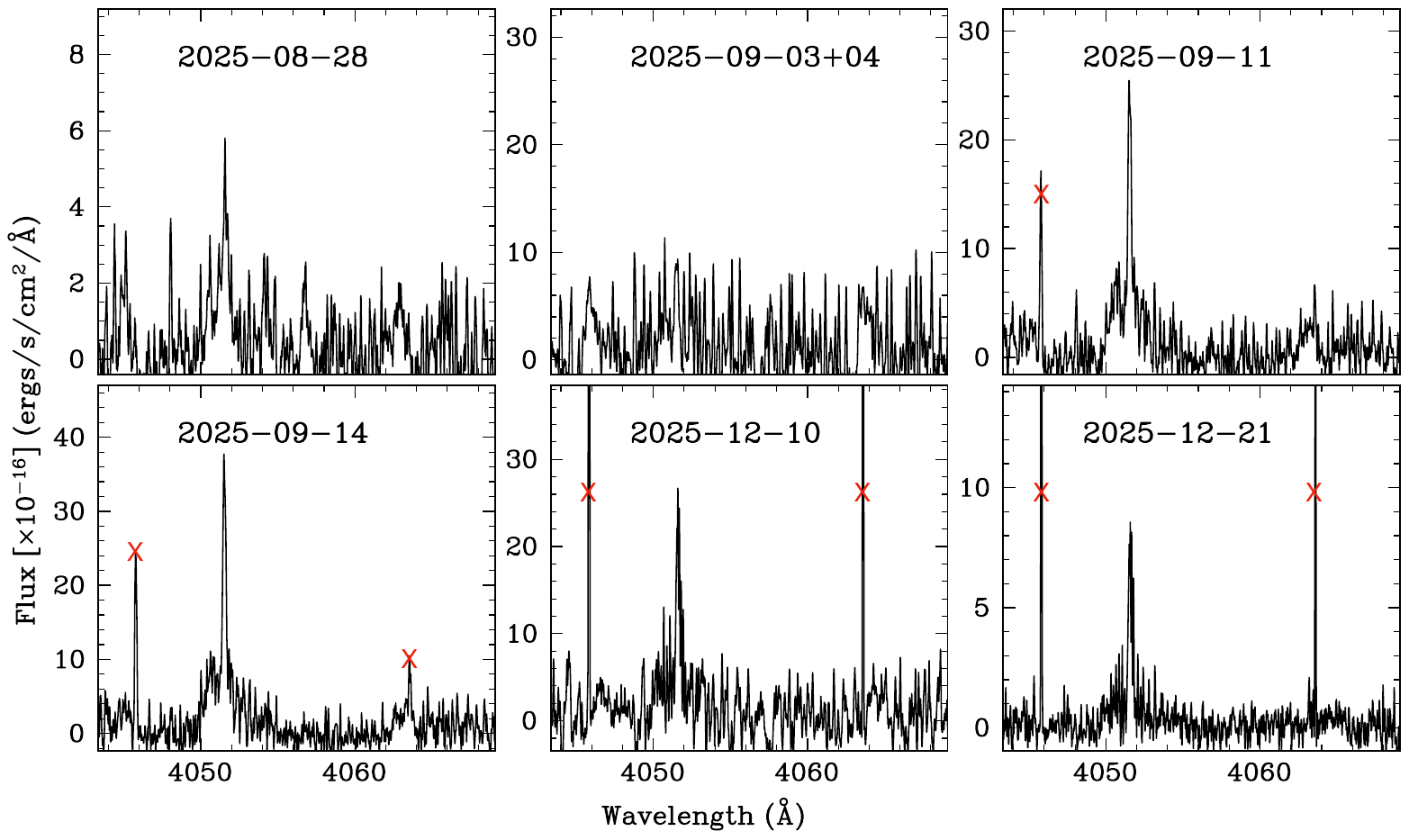}
\label{C3_panel}
\end{subfigure}\hfill
\caption{Evolution of the emissions C$_2$, CH and C$_3$, detected in 3I/ATLAS during the complete observing campaign. The red cross depicts NiI and FeI emission lines present in the region. \textit{Left}: C$_2$ Emissions detected in the UVES spectra on different epochs. \textit{Right}: CH Emissions detected in the UVES spectra on different epochs. \textit{Bottom}: C$_3$ Emissions detected in the UVES spectra on different epochs.}
\label{panel2}
\end{figure*}  

\newpage
\twocolumn
\section{Water from OH and [OI] line}\label{OH_O1}
\begin{figure}[h!]
  \centering
   \includegraphics[width=0.95\linewidth]{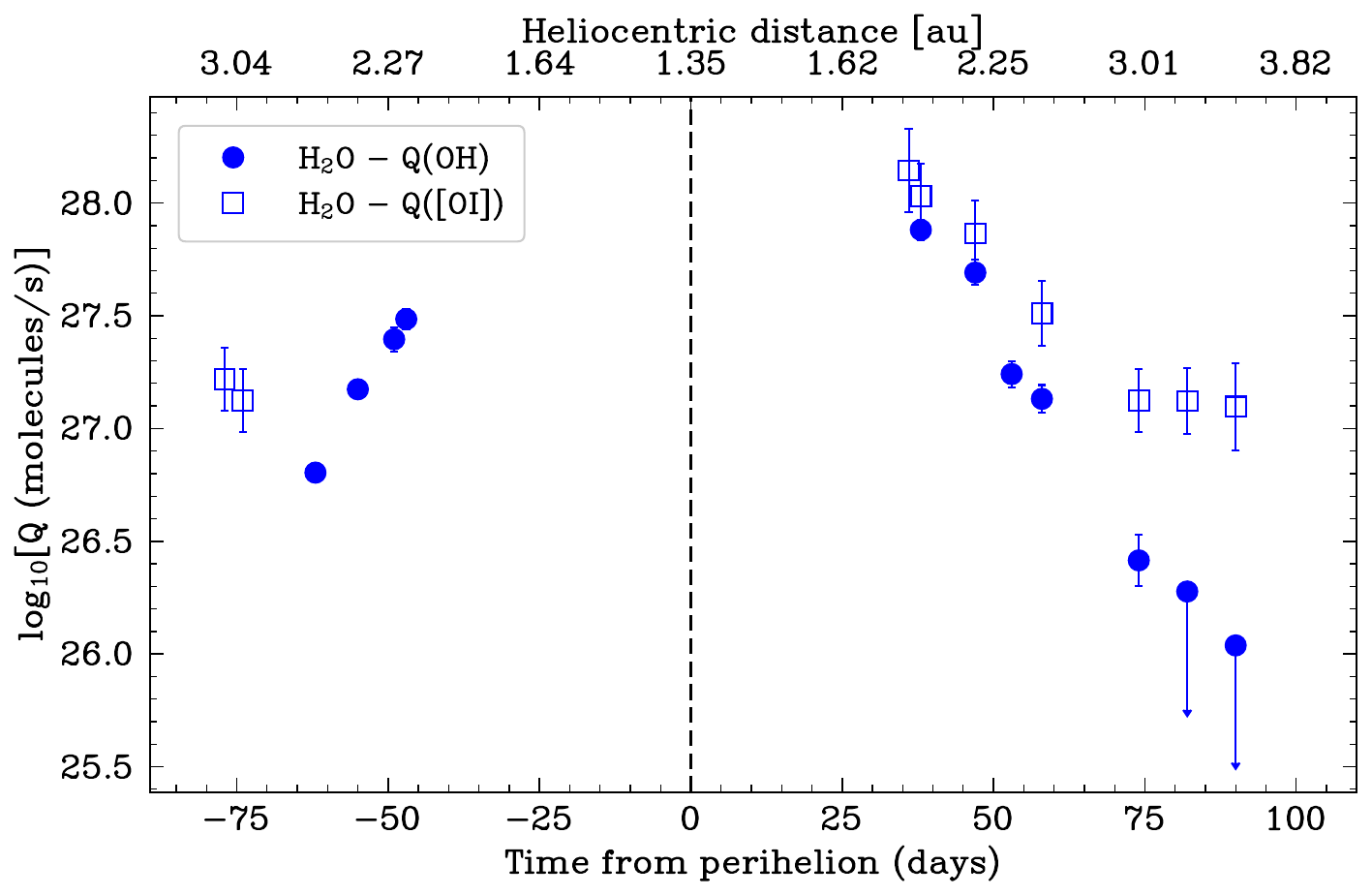}
      \caption{The water production rate of comet 3I/ATLAS computed from the OH emission and the forbidden oxygen red line.
              }
         \label{OHO1_water}
   \end{figure}

The comparison of OH-derived water production rates compared with that derived from the 6300 \AA~ [OI] line can be used to assess the source of origin of these red [OI] lines \citep[e.g.,][and references therein]{3I_kawakita2026}. We find that the [OI] derived water production values are systematically higher than the OH-based rates, although the two estimates tend to converge close to perihelion, as shown in Figure \ref{OHO1_water}. This behaviour can be explained because CO$_2$ also contribute to produce [OI] emission \citep{Bhardwaj2012, Raghuram2014}. 

Prior comet studies have shown that forbidden oxygen emission is sensitive to the relative roles of the H$_2$O and CO$_2$ in the coma \citep{festou81-oxygen}. The detection of [OI] emission prior to the appearance of OH provides independent support for a substantial CO$_2$ contribution to the oxygen budget. In 3I/ATLAS, [OI] lines are detected in the earliest pre-perihelion epochs while OH remains undetected, a behaviour also seen for comet C/2001 Q4 (NEAT) at 3.7 au \citep{decock13}. This suggests that CO$_2$, rather than H$_2$O alone, is a major parent of the observed oxygen emission. Hence, a persistent offset between [OI] and OH-derived water production points to a higher abundance of CO$_2$. The comparatively large abundance of CO$_2$ in 3I/ATLAS has indeed been reported by different studies using observations from the JWST \citep{JWST2025CO2,3I_JWST_cordiner_2026, 3I_JWST_Nathan}.

The fact that the two estimates are converging near perihelion suggests that water is the dominant driver of oxygen production when the comet was close to perihelion, while at other epochs CO$_2$ contribution likely elevated the oxygen-line-derived water proxy. This result ties the temporal evolution of the oxygen lines directly to the changing active volatile mix in the coma.

\section{3I/ATLAS compared with a dynamically new Solar System comet at similar geometries}\label{17K2_sec}
\begin{figure}[h!]
  \centering
   \includegraphics[width=0.95\linewidth]{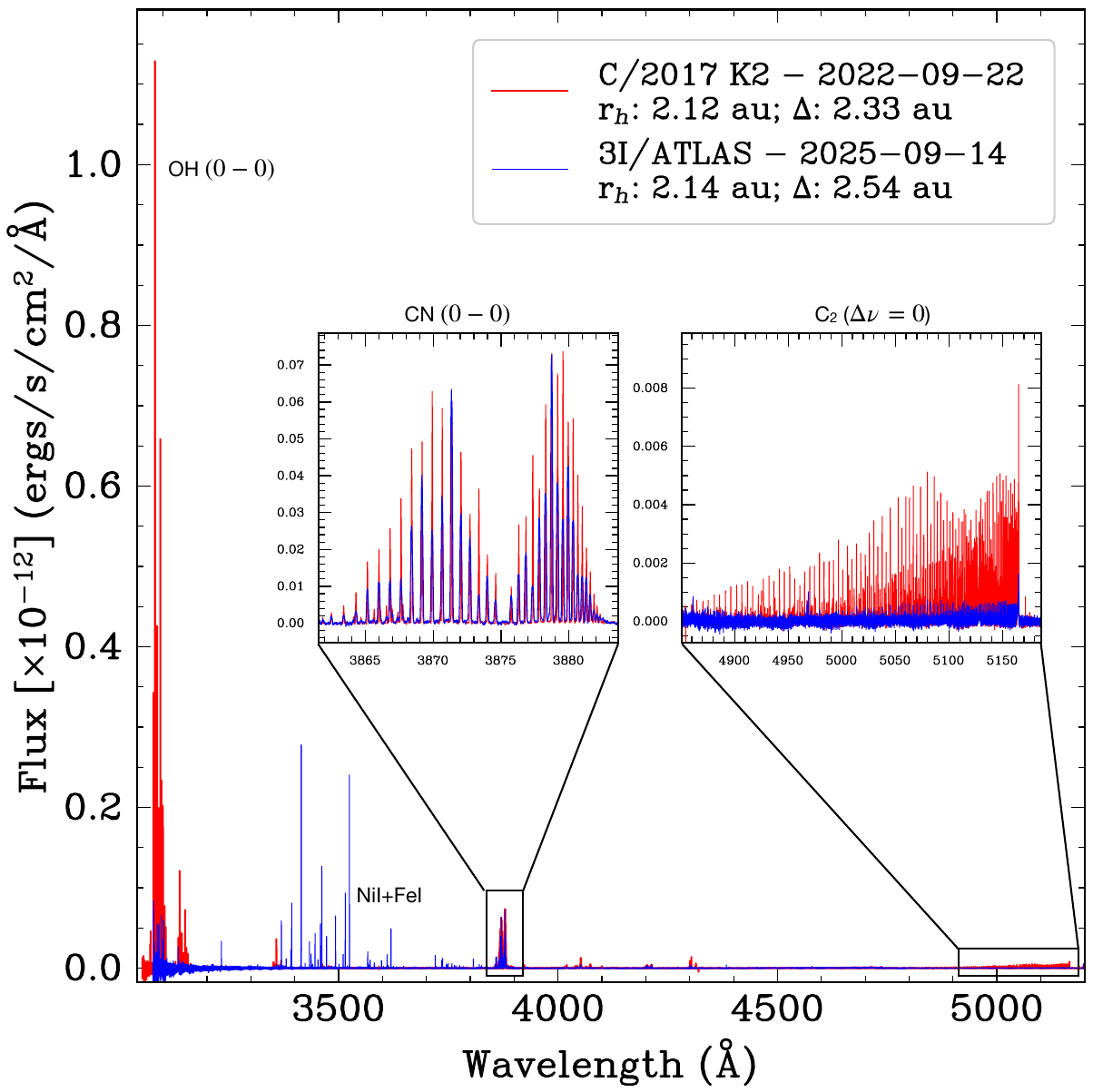}
      \caption{Comparison of UVES spectra obtained for C/2017 K2 \citep{Said_17K2} and 3I/ATLAS under similar geometry.}
         \label{3I_17K2}
   \end{figure}

We find an interesting comparison with the dynamically new comet C/2017 K2, observed with UVES \citep{Said_17K2} using the same instrumental setup and at comparable heliocentric and geocentric distances and phase angle. As shown in Figure~\ref{3I_17K2}, the flux-calibrated UVES spectrum of C/2017 K2 exhibits a CN band of visually comparable strength to that of 3I/ATLAS, whereas OH and C$_2$ emissions are markedly weaker in 3I/ATLAS, and metallic emission lines are noticeably stronger.

Although visual band intensities do not directly measure relative abundances due to heliocentric-velocity-dependent fluorescence efficiencies, the derived production rates provide a quantitative comparison. For 3I/ATLAS, we derive $Q_{\rm CN}=2.34\times10^{25}$~s$^{-1}$ and $Q_{\rm OH}=1.90\times10^{27}$~s$^{-1}$, compared with $Q_{\rm CN}=1.15\times10^{26}$~s$^{-1}$ and $Q_{\rm OH}=5.23\times10^{28}$~s$^{-1}$ for C/2017 K2 \citep{Said_17K2}. The resulting log$_{10}$(CN/OH) ratio is $-1.91$ for 3I/ATLAS, compared with $-2.69$ for C/2017 K2, corresponding to a CN/OH ratio approximately 5.6 times higher in 3I/ATLAS. This indicates that, at comparable heliocentric distance, 3I/ATLAS produces substantially less OH relative to CN than C/2017 K2. Since OH is the primary photodissociation product of H$_2$O, this result is consistent with comparatively lower water production relative to CN in 3I/ATLAS at this epoch. We do not interpret this as evidence that 3I/ATLAS is intrinsically water-poor, as absolute production rates depend on nucleus size, active fraction, and overall activity level. Rather, the comparison indicates a distinct relative production of OH and CN between the two comets.
\onecolumn
\section{3I/ATLAS versus other comets}
\begin{figure}[h!]
  \centering
   \includegraphics[width=0.92\linewidth]{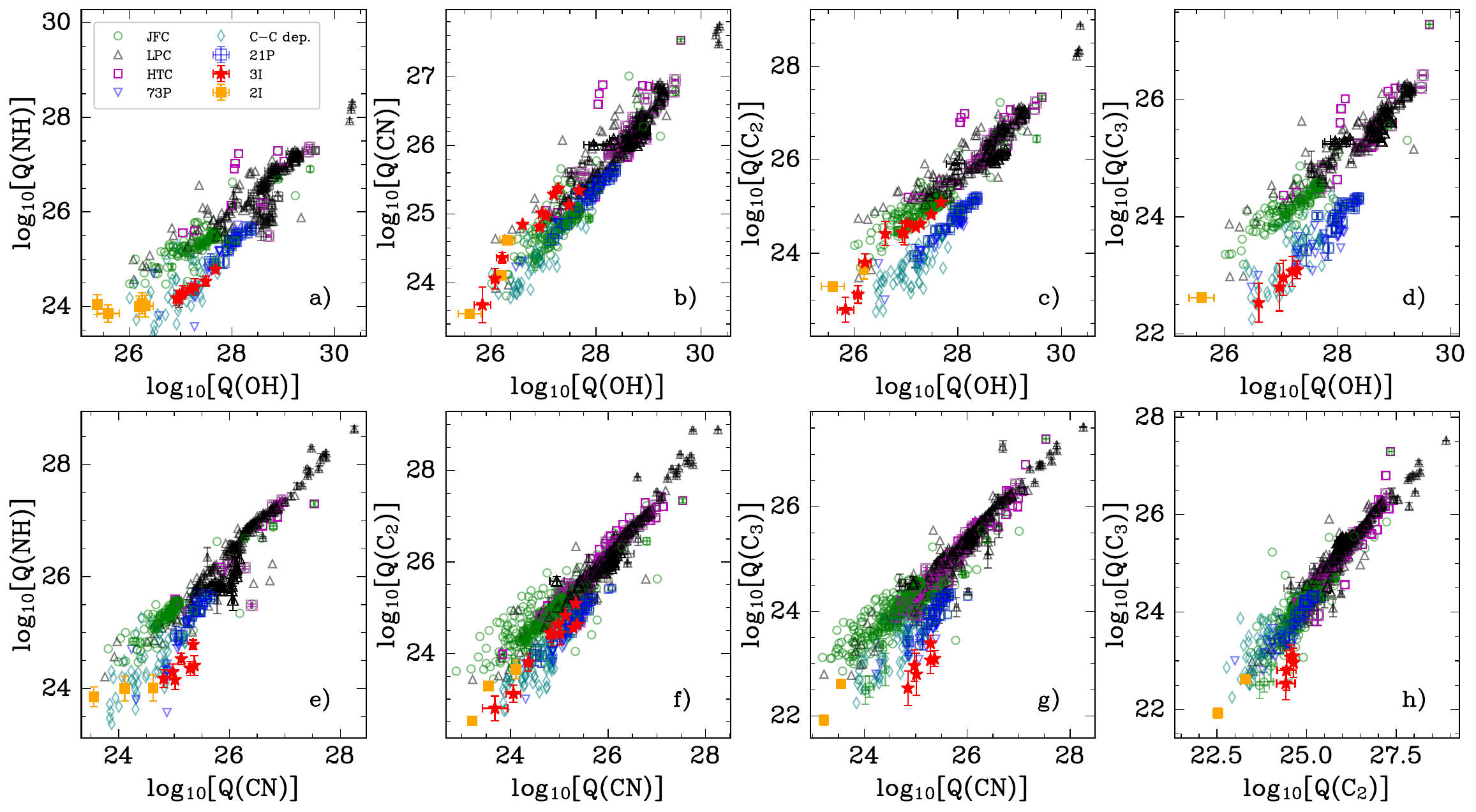}
      \caption{Comparison of the inter-dependence of the different molecular production rates in the case of 3I/ATLAS and a large sample of Jupiter Family Comets (JFC), Long period Comets (LPC) and Halley Type Comets (HTC) \citep{cochran_30years, Aravind_PhDT, opitom_C2012F6, opitom_C2013R1, moulane_46P, Said_17K2, 21P_moulane, Manfroid_Ni_Fe_comets, 73P_schleicher, mathieu_12P, JFC_cdepletion_schleicher, langland-shula} along with the first interstellar comet 2I/Borisov \citep{Borisov_highres, Bair_2I}. The symbols are given in the first upper left panel.}.
    \label{prod_rate_all_full}
   \end{figure}

\begin{figure}[h!]
  \centering
   \includegraphics[width=0.92\linewidth]{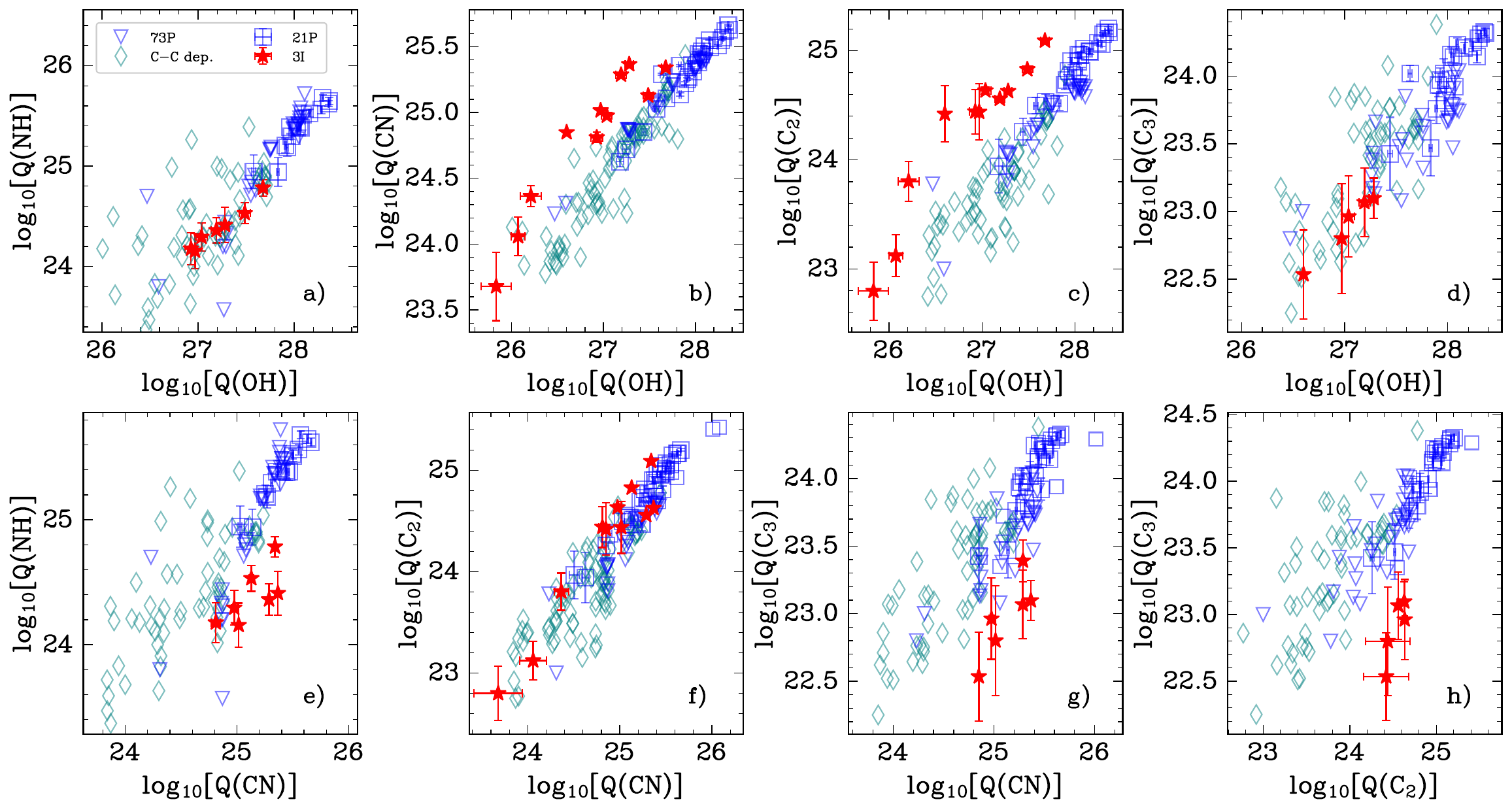}
      \caption{Comparison of the inter-dependence of the different molecular production rates in the case of 3I/ATLAS and a sample of strongly depleted comets \citep{21P_moulane, 21P_schleicher, 73P_schleicher, JFC_cdepletion_schleicher}. The symbols are given in the first upper left panel.}
         \label{stronglydepleted}
   \end{figure}
   

 \end{appendix}


\end{document}